\documentclass[journal]{IEEEtran}
\usepackage{amsmath,amsfonts}
\usepackage{amssymb}
\usepackage{bm}
\usepackage{relsize}
\usepackage{algorithmic}
\usepackage{algorithm}
\usepackage{array}
\usepackage[caption=false,font=normalsize,labelfont=sf,textfont=sf]{subfig}
\usepackage{textcomp}
\usepackage{stfloats}
\usepackage{url}
\usepackage{verbatim}
\usepackage{graphicx}
\usepackage{cite}
\usepackage{color}
\newtheorem{proposition}{\underline{Proposition}}
\newtheorem{corollary}{\underline{Corollary}}
\newtheorem{remark}{\underline{Remark}}
\begin{document}	
\title{MIMO Integrated Sensing and Communication Exploiting Prior Information}
\author{Chan Xu and Shuowen Zhang
\thanks{This paper was presented in part at the IEEE International Symposium on Information Theory (ISIT), Taipei, Taiwan, Jun. 2023 \cite{ISIT}.}
\thanks{The authors are with the Department of Electrical and Electronic Engineering, The Hong Kong Polytechnic University, Hong Kong SAR (e-mail: chan.xu@polyu.edu.hk; shuowen.zhang@polyu.edu.hk).}}
\maketitle
	
\begin{abstract}
In this paper, we study a multiple-input multiple-output (MIMO) integrated sensing and communication (ISAC) system where one multi-antenna base station (BS) sends information to a user with multiple antennas in the downlink and simultaneously senses the location parameter of a target based on its reflected echo signals received back at the BS receive antennas. We focus on the case where the location parameter to be sensed is \emph{unknown} and \emph{random}, for which the prior distribution information is available for exploitation. First, we propose to adopt the \emph{posterior Cram\'er-Rao bound (PCRB)} as the sensing performance metric with prior information, which quantifies a lower bound of the mean-squared error (MSE). Since the PCRB is in a complicated form, we derive a tight upper bound of it to draw more insights. Based on this, we analytically show that by exploiting the prior distribution information, the PCRB is always no larger than the CRB averaged over random location realizations without prior information exploitation. Next, we formulate the transmit covariance matrix optimization problem to minimize the sensing PCRB under a communication rate constraint. We obtain the \emph{optimal solution} and derive useful properties on its rank. Then, by considering the derived PCRB upper bound as the objective function, we propose a low-complexity suboptimal solution in \emph{semi-closed form}. Numerical results demonstrate the effectiveness of our proposed designs in MIMO ISAC exploiting prior information.
\end{abstract}
\begin{IEEEkeywords}
Integrated sensing and communication, multiple-input multiple-output (MIMO), posterior Cram\'er-Rao bound (PCRB).
\end{IEEEkeywords}
\section{Introduction}\label{sec_int}
Driven by new emerging sixth-generation (6G) applications such as vehicle-to-everything (V2X), platoons, virtual reality (VR), unmanned aerial vehicle (UAV), etc. \cite{6G}, there has been an ever-increasing demand for simultaneously providing high-capacity communication and high-precision sensing (or localization) services. Integrated sensing and communication (ISAC) \cite{ISAC_survey} is a promising technique to meet the aforementioned demands, where communication and sensing services share the same frequency band and are jointly designed. By accomplishing the communication and sensing functions over a single hardware unit and common wireless resources, ISAC systems are anticipated to save the hardware cost and energy consumption, and achieve enhanced sensing performance with proper wireless resource allocation. Particularly, in multi-antenna systems, smart antenna techniques in wireless communications and multiple-input multiple-output (MIMO) radar techniques in sensing share similar spirits, which makes them perfect platforms to realize dual-function radar-communication (DFRC) \cite{ISAC_survey2}.
	
To fully reap the benefits of ISAC, it is of paramount importance to optimally design the transmit signals to accomplish dual functionalities of target sensing and communication, which has recently attracted significant research attention. The ISAC transmit signal designs can be generally divided into three categories: radar-centric design, communication-centric design, and joint design. In radar-centric design, the communication information is embedded in the common radar pulse by utilizing its extra potential information or parameters, such as the rate and phase parameters of chirp signals \cite{signal_design1}, the distinct level of the sidelobe \cite{signal_design2}, as well as the antenna and subcarrier indices \cite{signal_design3}. In communication-centric design, the classic communication signals are modified to achieve the sensing functionality. Orthogonal frequency division multiplexing (OFDM) has been widely used in current communication systems and can be employed in radar with minimal changes to signal processing \cite{signal_design4,Shi,Liu,Wang,Shi2}. By exploiting the Doppler sensitivity arising from the multicarrier structure of the OFDM signal, the Doppler ambiguity of the pulsed Doppler radars can be solved \cite{signal_design5}. However, in radar-centric design, the communication rate is limited by the pulse repetition frequency of radar; in communication-centric design, the sensing performance is restricted since the communication waveform is not well shaped to satisfy the sensing-specific constraints. To provide a better trade-off between communication and sensing, the transmit signal should be designed with joint consideration of both functions.
	
In the existing literature on dual-functional ISAC transmit signal design, the rate or signal-to-interference-plus-noise ratio (SINR) is typically used to characterize the communication performance. The sensing performance metrics can be generally categorized into three classes: radar beampattern approximation, radar SINR, and estimation-theoretic metric. In the first class, the transmit signals are designed to approximate a desired and pre-designed radar beampattern \cite{jointDesign1,jointDesign2}. In the second class, the transmit signals are designed to guarantee the SINR of the echo signal \cite{jointDesign3,jointDesign4}. Based on these two metrics, the sensing performance is implicitly reflected by the difference between the desired beampattern and the approximated beampattern or the SINR, and cannot be explicitly quantified. In the third class, the mean-squared error (MSE) is a commonly adopted metric to assess the sensing performance. However, since the minimum possible MSE is generally difficult to characterize, some lower bounds of the MSE have been proposed, among which the most well-known one is the \textit{Cram\'{e}r-Rao bound (CRB)} \cite{CRB1}. For MIMO radar systems, the expressions of CRB for angle estimation \cite{CRB2} and velocity estimation \cite{CRB3} have been derived. With CRB as the sensing performance metric, various works have studied the joint transmit signal design \cite{jointDesign5,jointDesign6,jointDesign7}.
	
Note that CRB is a function of the exact values of the parameters to be sensed (estimated), which is only suitable for \emph{deterministic} parameters, and the transmit signal can only be designed based on the CRB corresponding to a \emph{given parameter (e.g., location)}. However, in practice, the parameters to be sensed can be \emph{unknown} and \emph{random}, while the parameter distribution can be known \textit{a priori}. For example, for a mobile vehicle or pedestrian, the location parameters at the upcoming time slots are generally functions of the location parameters at the current and previous time slots, for which the distributions can be obtained based on the previous localization results and/or by exploiting empirical data. By employing the \textit{a priori} probability function of the random parameter, Van Trees presented a lower bound that is analogous to the CRB for MSE and suitable for random parameters \cite{PCRB1}, which is known as \textit{posterior Cram\'{e}r-Rao bound (PCRB)}\footnote{It is worth noting that there exists another related concept termed prior CRB \cite{Prior_CRB}. In the context of (posterior) PCRB, the \textit{a priori} knowledge is the distribution of the random parameters; while in the context of prior CRB, the exact values of a subset of the parameters are known \textit{a priori}.} \cite{PCRB2} or Bayesian CRB (BCRB) \cite{BCRB}. Note that in contrast to CRB, PCRB is \emph{not} dependent on the exact values of the parameters to be sensed.
		
In the context of radar, PCRB has been employed in target tracking by leveraging the previous localization results. Target tracking is usually modeled as a discrete-time process, where the estimation in the current time slot depends on not only the observations up to now but also the evolution of states. In a discrete-time system, PCRB can be derived from the joint distribution of the observations and the states up to the current time \cite{PCRB3}. Specifically, based on the fact that the previous observations are already known, the conditional PCRB can be derived from the posterior probability density function (PDF) that is conditioned on the past observations up to now \cite{PCRB4,Active}. With conditional PCRB as the performance metric, the transmit signals of MIMO radar can be adaptively optimized at each time slot with updated posterior PDF \cite{detection1,detection2}. Recently, the sequential transmit signal design has been employed in ISAC systems to achieve a better trade-off between communication performance and sensing performance \cite{ISAC_PCRB1,ISAC_PCRB2}. In \cite{ISAC_PCRB1}, based on the previous observations, the posterior PDFs of random parameters at the current time were updated, based on which the signal power towards the possible target locations was maximized. In \cite{ISAC_PCRB2}, the transmit signal was optimized to minimize the PCRB in sensing the user's location, which was updated at each time slot. The existing works that employ the prior information are based on the assumption that the distribution function is unknown and inferred according to previous observations and estimations. Due to the unavoidable error in estimation, the inferred distribution function may be inexact, and it will take several attempts to obtain satisfactory sensing MSE.
	
It is worth noting that the PDF of various sensing parameters in practice can be directly obtained based on empirical data or statistical analysis. For example, the probability of appearance of a target at each geographical location can be obtained based on historic observations and empirical data; the random movement of a target caused by wind or turbulence can be typically characterized by a random walk model \cite{random2}. How to exploit such prior distribution information for designing the transmit signals in ISAC systems to achieve an optimal trade-off between the PCRB and communication rate still remains an open problem. Along this line, in our prior work \cite{ISIT}, we considered a MIMO radar system, and studied the transmit signal optimization to minimize the PCRB exploiting prior distribution information; in another prior work \cite{SecureISAC}, we studied the transmit signal optimization in a secure multiple-input single-output (MISO) ISAC system. However, for a general MIMO ISAC system, finding the optimal transmit signal design remains an unaddressed problem.
	
\begin{figure}
		\centering
		\includegraphics[width=3.5in]{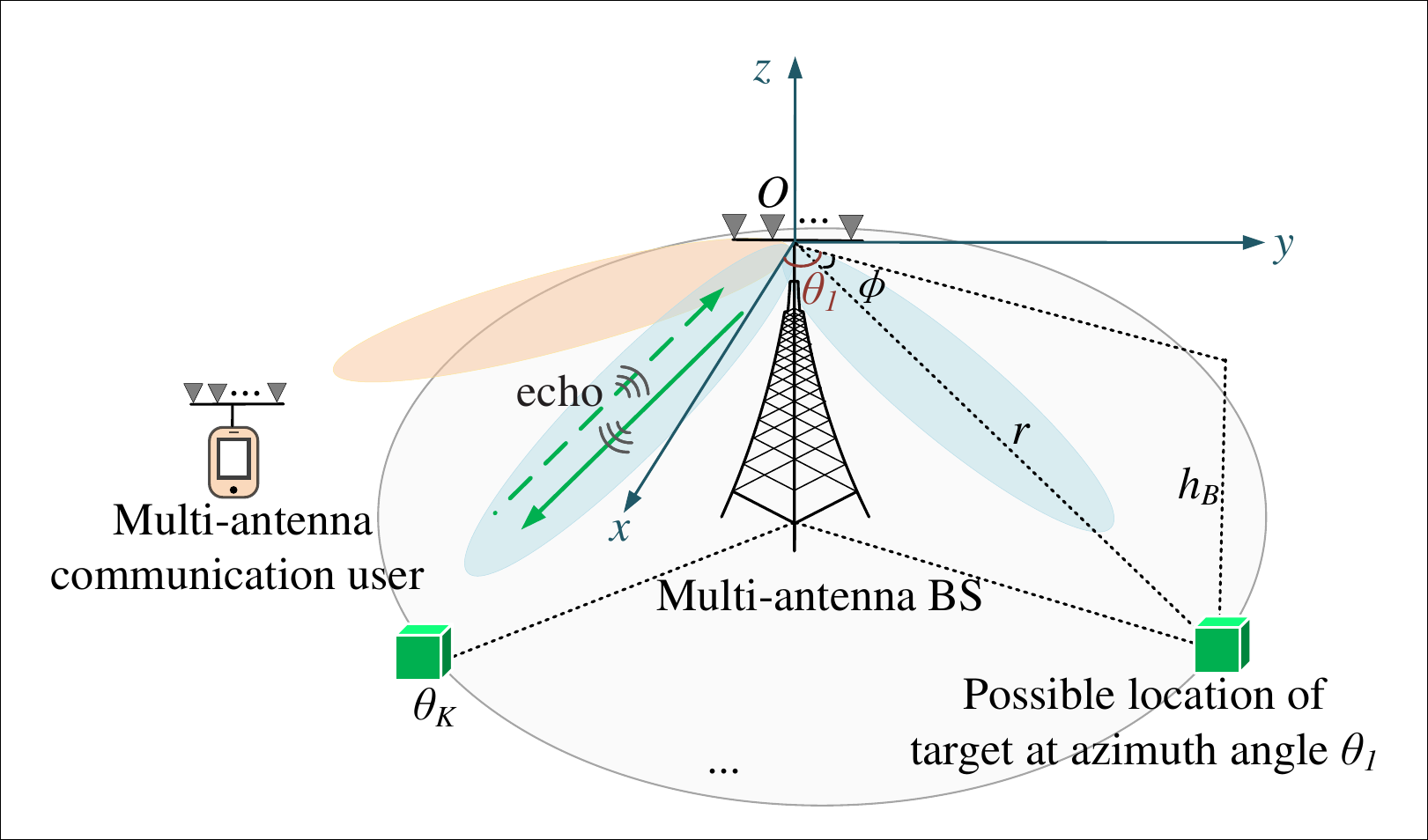}
		\vspace{-6mm}
		\caption{Illustration of a MIMO ISAC system with unknown and random target location.}\label{fig_sys}
		\vspace{-6mm}
\end{figure}
	
In this paper, we consider a MIMO ISAC system where a multi-antenna BS sends dual-function signals to communicate with a multi-antenna user and sense the \emph{unknown} and \emph{random} location parameter of a target, as illustrated in Fig. \ref{fig_sys}. Specifically, sensing is performed based on the echo signals reflected by the target and arrive back at the BS receive antennas. The prior PDF of the location parameter is known for exploitation. Our main contributions are summarized as follows.
\begin{itemize}
	\item {First, to characterize the sensing performance exploiting prior distribution information, we characterize the PCRB of the MSE, which is in a complicated form. We then derive a tractable and tight upper bound for the PCRB. Based on this, we analytically show that by exploiting prior distribution information, the PCRB is always no larger than the average CRB without exploiting the prior information.}
	\item {Next, we formulate an optimization problem for the transmit covariance matrix to minimize the sensing PCRB, subject to a communication rate constraint for the user. The formulated problem is challenging to solve due to the complicated fractional expression of the PCRB. To tackle this problem, we first transform it into an equivalent convex problem via leveraging Schur complement, based on which the optimal solution is obtained. By applying the Lagrange duality method, we obtain useful insights on the optimal rank of the transmit covariance matrix with both sensing and MIMO communication functions. We then propose a suboptimal solution with lower complexity by minimizing the upper bound of the PCRB, which has a semi-closed form.}
	\item {Finally, the performance and complexity of the proposed transmit covariance matrix designs are evaluated via numerical examples. The efficacy of the proposed PCRB upper bound is validated numerically. Moreover, the suboptimal solution is observed to achieve close performance to the optimal solution under low-to-moderate number of transmit antennas. It is also shown that the sensing performance achieved by our proposed designs outperforms that achieved by a benchmark scheme based on known but inexact target location, and is close to that achieved by a genie-aided scheme with exactly known target location, thanks to the smart exploitation of prior distribution information.}
\end{itemize}
	
The remainder of this paper is organized as follows. Section \ref{sec_sys} presents the system model. Section \ref{sec_PCRB} characterizes the sensing performance based on PCRB. The transmit covariance matrix optimization problem is formulated to minimize the sensing PCRB in Section \ref{sec_pro}. For the formulated problem, Section \ref{sec_sol} presents both the optimal and suboptimal solutions, and draws useful insights on the rank of the optimal transmit covariance matrix. Numerical results and their pertinent discussions are presented in Section \ref{sec_num}. Finally, Section \ref{sec_con} concludes this paper.
	
\textit{Notations:} Vectors and matrices are denoted by boldface lower-case letters and boldface upper-case letters, respectively. $\mathbb{C}^{N\times L}$ and $\mathbb{R}^{N\times L}$ denote the space of $N\times L$ complex matrices and the space of $N\times L$ real matrices, respectively. $\bm{I}_N$ denotes an $N\times N$ identity matrix, and $\bm{0}$ denotes an all-zero matrix with appropriate dimension. For a square matrix $\bm{W}$, $\mathrm{tr}(\bm{W})$, $|\bm{W}|$, and $ \bm{W}^{-1}$ denote its trace,  determinant, and inverse, respectively. $\bm{W}\succeq \bm{0}$ and $\bm{W}\succ \bm{0}$ mean that $\bm{W}$ is positive semi-definite and positive definite, respectively. For an $N\times L$ matrix $\bm{H}$, $\bm{H}^H$, $\mathrm{rank}(\bm{H})$, $ \|\bm{H}\|_F $, and $[\bm{H}]_{i,j}$ denote its conjugate transpose, rank, Frobenius norm, and $(i,j)$-th element, respectively. $\mathrm{diag}\{x_1,...,x_N\}$ denotes an $N\times N$ diagonal matrix with $x_1,...,x_N$ being the diagonal elements. $\|\bm{x}\|$ denotes the $l_2$-norm of a complex vector $\bm{x}$. $x^*$ and $|x|$ denote the conjugate and absolute value of a complex scalar $x$, respectively. $(z)^+=\mathrm{max}(z,0)$ with $\mathrm{max}(a,b)$ denoting the maximum between two real numbers. The distribution of a circularly symmetric complex Gaussian (CSCG) random variable with mean $\mu$ and variance $\sigma^2$ is denoted by $\mathcal{CN}(\mu,\sigma^2)$, and $\sim$ means ``distributed as''. $\mathbb{E}_\theta[\cdot]$ denotes the statistical expectation over parameter $\theta$. $\mathcal{O}(\cdot)$ denotes the standard big-O notation.
	
\section{System Model}\label{sec_sys}	
We consider a MIMO ISAC system, which consists of a BS equipped with $N_t\geq 1$ transmit antennas and $N_r\geq 1$ co-located receive antennas, a communication user equipped with $N_u\geq1$ receive antennas, and a point target with an \emph{unknown} and \emph{random} location, as illustrated in Fig. \ref{fig_sys}. The distribution of the target's location information is assumed to be known \emph{a priori} based on empirical data or target movement pattern. The BS aims to deliver information to the user in the downlink and estimate the location parameter of the target by exploiting its prior distribution information. Specifically, we consider a three-dimensional (3D) spherical coordinate system with a reference point at the BS antenna arrays being the origin, as illustrated in Fig. \ref{fig_sys}. We assume that the target is located on the ground and the height of the BS's antennas is $h_B\geq 0$ in meters (m). For the purpose of drawing essential insights, we assume that every possible target location has the same distance $r\geq 0$ m, the same elevation angle $\phi \in [-\frac{\pi}{2}, \frac{\pi}{2})$ and a different azimuth angle with respect to the origin, where the common distance (range) information $r$ is known \emph{a priori}.\footnote{The range information can be either known as prior information of the target, or estimated efficiently via e.g., time-of-arrival (ToA) methods.} Note that the elevation angle of the target is consequently known as $\phi=\arcsin \frac{ -h_B}{r}$. Thus, we focus on the estimation of the only \emph{unknown} and \emph{random} parameter, i.e., the target's azimuth angle denoted by $\theta\in[-\frac{\pi}{2}, \frac{\pi}{2})$. The PDF of $\theta$ is denoted by $p_\Theta(\theta)$, which is known at the BS.
	
Consider a quasi-static block-fading channel model between the BS and the user, where the channel remains constant within each coherence block consisting of $L_c$ symbol intervals, and may vary independently over different blocks. We consider a narrowband system and let $\bm{H} \in \mathbb{C}^{N_u\times N_t}$ denote the channel matrix from the BS to the user in the coherence block of interest, which is assumed to be known perfectly at both the BS and the user. On the other hand, let $L$ denote the number of symbol intervals used for the estimation of $\theta$, within which $\theta$ remains unchanged. In this paper, we focus on the case of $L\leq L_c$, and study the dual-function transmit signal optimization within $L$ symbol intervals.\footnote{It is worth noting that our results can also be directly applied to the case of $L>L_c$, by designing the dual-function transmit signals in each channel coherence block separately.}
	
Let $\bm{x}_l\in \mathbb{C}^{N_t\times 1}$ denote the baseband equivalent dual-function transmitted signal vector in the $l$-th symbol interval, $l=1,...,L$. The received signal at the user receiver in each $l$-th symbol interval is given by
\begin{align}
	\bm{y}_l^C = \bm{H}\bm{x}_l+ \bm{n}_l^C ,\quad  l=1,...,L,
\end{align}
where $\bm{n}_l^C \sim \mathcal{CN}(\bm{0},\sigma_c^2\bm{I}_{N_u})$  denotes the CSCG noise vector over the user's receive antennas in the $l$-th symbol interval, with $\sigma_c^2$ denoting the average noise power. Specifically, $\bm{x}_l$ represents the linearly precoded information symbols to be delivered to the user in the $l$-th symbol interval, where a constant transmit covariance matrix denoted by $\bm{W}=\mathbb{E}[\bm{x}_l\bm{x}_l^H],\ \forall l$ is applied to all symbol intervals since both the communication channel and the target's location to be estimated remain unchanged  within $L$ symbol intervals. We consider a sum transmit power constraint denoted by $\mathrm{tr}(\bm{W})\leq P$, where $P$ denotes the total power budget. The achievable communication rate for the user is thus expressed as
\begin{align}\label{capacity}
	R =\log_2\left|\bm{I}_{N_u}+ \frac{\bm{H} \bm{W}\bm{H}^H}{\sigma_c^2}\right| 	
\end{align}
in bits per second per Hertz (bps/Hz).
	
Besides reaching the user receiver, the transmitted signals will be reflected by the target back to the BS receiver; $\theta$ is then estimated by processing the received echo signals. We consider a line-of-sight (LoS) channel model between the target and the BS where no obstruction/scatter exists between the BS transceiver and each possible target location.\footnote{Note that in device-free sensing, the target is typically located in the LoS range of the BS responsible for sensing, otherwise the reflected echo signal will be too weak under the round-trip channel. Moreover, if additional multi-path components exist in the channel, their effect on the round-trip reflected echo signals is generally small and can be treated as additional noise. On the other hand, the communication user may be located farther away from the BS. Thus, we consider that $\bm{H}$ consists of both the LoS component and the non-LoS (NLoS) component with independent and identically distributed (i.i.d.) entries.} The overall channel from the BS transmitter to the BS receiver via target reflection is given by
\begin{align}
	\bm{G}(\theta)=\bm{h}_R(\theta)\psi\bm{h}^H_T(\theta).	
\end{align}
Specifically, $\psi\in \mathbb{C}$ denotes the radar cross-section (RCS) coefficient, which is an unknown and deterministic parameter. $\bm{h}_R(\theta)=\frac{\sqrt{\beta_0}}{r}\bm{b}(\theta)$ and $\bm{h}^H_T(\theta)=\frac{\sqrt{\beta_0}}{r}\bm{a}^H(\theta)$ denote the target-receiver and transmitter-target channel vectors, respectively, where $\beta_0$ denotes the reference channel power at reference distance $1$ m; $\bm{a}^H(\theta)$ and $\bm{b}(\theta)$ denote the transmit/receive antenna array steering vectors given by
\begin{align}
	&a_n(\theta)=e^{\frac{-j\pi d(N_t-2n+1)\cos\phi\sin\theta}{\lambda}},\quad n=1,...,N_t,\\
	&b_m(\theta)=e^{\frac{-j\pi d(N_r-2m+1)\cos\phi\sin\theta}{\lambda}},\quad m=1,...,N_r,
\end{align}
with $\cos\phi=\frac{\sqrt{r^2-h_B^2}}{r}$, $d$ denoting the antenna spacing in m, and $\lambda$ denoting the wavelength in m. For simplicity, we define $\alpha\overset{\Delta}{=}\frac{\beta_0}{r^2}\psi=\alpha_R+j\alpha_I$ as the overall reflection coefficient containing both the round-trip path loss and target reflection, which yields $\bm{G}(\theta)=\alpha \bm{b}(\theta)\bm{a}^H(\theta)$. It is worth noting that in general, $\alpha$ is an \emph{unknown} and \emph{deterministic} parameter.

Hence, the received echo signal vector at the BS receive antennas in the $l$-th symbol interval is given by
\begin{align}
	{\bm{y}_l^S}=\bm{G}(\theta)\bm{x}_l+{\bm{n}_l^S},\quad l=1,...,L,
\end{align}
where $ \bm{n}_l^S \sim\mathcal{CN}(\bm{0},\sigma_s^2\bm{I}_{N_r})$  denotes the CSCG noise vector over the BS receive antennas in the $l$-th symbol interval, with $\sigma_s^2$ denoting the average noise power. The collection of received signal vectors at the BS receiver over $L$ symbol intervals is thus given by
\begin{align}\label{Y}
	\bm{Y}=[{\bm{y}_1^S},...,{\bm{y}_L^S} ]=\bm{G}(\theta)\bm{X}+[{\bm{n}_1^S},...,{\bm{n}_L^S} ],
\end{align}
where $\bm{X}=[\bm{x}_1,...,\bm{x}_L]$ denotes the collection of the transmitted signal vectors over $L$ symbol intervals.

Note that the received echo signals in $\bm{Y}$ and consequently the performance of estimating $\theta$ are critically determined by the transmit signal design, particularly for the case considered in this paper where prior distribution information about $\theta$ is available for exploitation. For example, to optimize the sensing performance, the radiated signal power should be more concentrated over the possible target angles with high probability densities. On the other hand, the transmit signal design also affects the communication rate $R$. To achieve a high rate, the transmit covariance matrix $\bm{W}$ needs to cater to the spatial sub-channels in $\bm{H}$. Therefore, with the limited spatial and power resources at the transmitter, there exists a non-trivial trade-off between the sensing performance and the communication performance. To resolve this trade-off, we will first establish a PCRB-based framework for characterizing the sensing performance in Section \ref{sec_PCRB}; then, we will study the transmit signal optimization towards the optimal balance between the sensing performance and communication performance in Sections \ref{sec_pro} and \ref{sec_sol}.

\section{Sensing Performance Characterization via PCRB}\label{sec_PCRB}
Conventionally, CRB has been widely adopted to characterize the estimation performance of unknown deterministic parameters, which is a lower bound of the MSE. In this section, by exploiting the prior distribution information of the unknown random parameter $\theta$, i.e., $p_\Theta(\theta)$, we propose to derive the PCRB of the MSE as the sensing performance metric.

\subsection{Derivation of PCRB}
We aim to estimate $\theta$ from the received signal $\bm{Y}$ shown in (\ref{Y}), which is a function of both the unknown random parameter $\theta$ and the unknown deterministic parameter $\alpha$. Hence, $\alpha$ needs to be jointly estimated with $\theta$ to obtain an accurate estimation of $\theta$. For ease of exposition, we define $\bm{\zeta}=[\theta,\alpha_R,\alpha_I]^T$ as the collection of all the unknown parameters.

The joint distribution of the observation $\bm{Y}$ and unknown parameter $\bm{\zeta}$ can be expressed as
\begin{equation}\label{jointpdf}
	f(\bm{Y},\bm{\zeta})=f(\bm{Y}|\bm{\zeta})p_Z(\bm{\zeta}),
\end{equation}
where $f(\bm{Y}|\boldsymbol{\zeta})$ is the conditional PDF of $\bm{Y}$ given $\bm{\zeta}$; $p_Z(\bm{\zeta})$ denotes the marginal distribution of $\boldsymbol{\zeta}$.

Note that since $\boldsymbol{\zeta}$ consists of a random parameter $\theta$ for which the distribution is known, the information of $\boldsymbol{\zeta}$ can be extracted by jointly exploiting the conditional PDF $f(\bm{Y}|\boldsymbol{\zeta})$ of the observation $\bm{Y}$ and the prior information of $\theta$. Specifically, based on (\ref{jointpdf}), the Fisher information matrix (FIM) for estimating $\boldsymbol{\zeta}$ is given by \cite{shen}
\begin{equation}\label{FIM}
	\bm{F} = \bm{F}_o +\bm{F}_p,
\end{equation}
where $\bm{F}_o$ represents the FIM from observation given as
\begin{equation}\label{Fo}
	\bm{F}_o=\mathbb{E}_{\bm{Y},\boldsymbol{\zeta} }\left[\frac{\partial \ln(f(\bm{Y}|\boldsymbol{\zeta}))}{\partial \boldsymbol{\zeta} }\left(\frac{\partial \ln(f(\bm{Y}|\boldsymbol{\zeta}))}{\partial \boldsymbol{\zeta} }\right)^H\right];
\end{equation}
$\bm{F}_p$ represents the FIM from prior information given as
\begin{equation}\label{Fp}
	\bm{F}_p=\mathbb{E}_{\boldsymbol{\zeta} }\left[\frac{\partial \ln(p_Z(\boldsymbol{\zeta} ))}{\partial \boldsymbol{\zeta} }\left(\frac{\partial \ln(p_Z(\boldsymbol{\zeta} ))}{\partial \boldsymbol{\zeta} }\right)^H\right].
\end{equation}

In the following, we derive more tractable expressions of the FIMs in (\ref{Fo}) and (\ref{Fp}). First, for $\bm{F}_o$, the log-likelihood function for estimating $\bm{\zeta}$ from the observation $\bm{Y}$ is expressed as \cite{sonars}:
\begin{align}
	\ln(f(\bm{Y}|\boldsymbol{\zeta}))&=\frac{2}{\sigma_s^2}\mathrm{Re}\{ \mathrm{tr}(\bm{X}^H\bm{G}^H(\theta)\bm{Y })\}\nonumber\\
	&-\frac{\|\bm{Y }\|_F^2 + \|\bm{G}(\theta)\bm{X}\|_F^2 }{\sigma_s^2}-N_rL\ln(\pi\sigma_s^2).
\end{align}
Since $\bm{G}(\theta)$ is a function of $\bm{a}(\theta)$ and $\bm{b}(\theta)$, $\bm{F}_o$ is a function of the derivatives of $\bm{a}(\theta)$ and $\bm{b}(\theta)$ denoted by $\dot{\bm{a}}(\theta)$ and $\dot{\bm{b}}(\theta)$, respectively, with $\dot{a}_n(\theta)=\frac{-j\pi d(N_t-2n+1)}{\lambda}\cos\phi\cos\theta a_n(\theta), n=1,...,N_t$ and $\dot{b}_m(\theta)=\frac{-j\pi d(N_r-2m+1)}{\lambda}\cos\phi\cos\theta b_m(\theta), m=1,...,N_r$. The FIM $\bm{F}_o$ in (\ref{Fo}) can be partitioned as
\begin{align}
	\bm{F}_o
	= \left[
	\begin{array}{ll}
		J_{\theta\theta}            & \bm{J}_{\theta\alpha}     \\
		\bm{J}_{\theta\alpha}^H & \bm{J}_{\alpha\alpha}
	\end{array}
	\right].
\end{align}
By noting that $\boldsymbol{a}^H(\theta)\dot{\bm{a}}(\theta)=0$ and $\boldsymbol{b}^H(\theta)\dot{\bm{b}}(\theta)=0$, $J_{\theta\theta}$, $\bm{J}_{\theta\alpha}$, and $\bm{J}_{\alpha\alpha}$ are given by \cite{sonars}
\begin{align} 		
	J_{\theta\theta}=&\frac{2|\alpha|^2L}{\sigma_s^2}\mathrm{tr}\left(\bm{A}_1\bm{W}\right)+\frac{2|\alpha|^2L }{\sigma_s^2}\mathrm{tr}\left(\bm{A}_2\bm{W}\right),  \\
	\bm{J}_{\theta \alpha}=& \frac{2L  }{\sigma_s^2} \mathrm{tr}\left(\bm{A}_3\bm{W}\right)[\alpha_R,\alpha_I],\\
	\bm{J}_{\alpha\alpha}=& \frac{2L  }{\sigma_s^2}\mathrm{tr}\left(\bm{A}_4\bm{W}\right)  \bm{I}_{2},
\end{align}
where
\begin{align}
	\bm{A}_1&=\int \|\dot{\bm{b}}(\theta)\|^2\bm{a}(\theta)\bm{a}^H(\theta) p_\Theta(\theta)d\theta, \\
	\bm{A}_2&=N_r\int \dot{\bm{a}}(\theta)\dot{\bm{a}}^H(\theta) p_\Theta(\theta)d\theta,\\
	\bm{A}_3&=N_r\int \dot{\bm{a}}(\theta)\bm{a}^H(\theta) p_\Theta(\theta)d\theta,\\
	\bm{A}_4&=N_r\int  \bm{a}(\theta)\bm{a}^H(\theta) p_\Theta(\theta)d\theta.
\end{align}
On the other hand, for $\bm{F}_p$, since $\alpha_R$ and $\alpha_I$ are deterministic parameters, we have $\frac{\partial
	\ln(p_Z(\boldsymbol{\zeta} ))}{\partial \boldsymbol{\zeta} }=\left[ \frac{\partial \ln(p_\Theta(\theta))}{\partial \theta},0,0\right]^T$, which yields $[\bm{F}_p]_{1,1}=\mathbb{E}_\theta\left[\left(\frac{\partial \ln(p_\Theta(\theta))}{\partial \theta}\right)^2\right]$ and $[\bm{F}_p]_{m,n}=0$ for any $(m,n)\neq (1,1)$.

Based on $\bm{F}$ given above, the overall PCRB for the MSE of estimating $\bm{\zeta}$ is determined by $\bm{F}^{-1}$, and the PCRB for estimating the desired sensing parameter $\theta$ is given by $\mathrm{PCRB}_{\theta}=[\bm{F}^{-1}]_{1,1}$ \cite{PCRB1}.

It is worth noting that the presented PCRB framework is general for characterizing a lower bound of the sensing MSE under any distribution $p_\Theta(\theta)$. In this paper, for the purpose of exposition, we will derive the PCRB corresponding to a practical PDF in an explicit manner, based on which the transmit covariance matrix will be optimized. Specifically, motivated by practical scenarios where the target's azimuth angle distribution is typically concentrated around one or multiple nominal azimuth angles, we assume that the PDF of $\theta$ follows a Gaussian mixture model, which is the weighted summation of $K\geq 1$ Gaussian PDFs, with each $k$-th Gaussian PDF having mean $\theta_k\in [-\frac{\pi}{2}, \frac{\pi}{2})$, variance $\sigma_k^2$, and carrying a weight of $p_k\in [0,1]$ that satisfies $\sum_{k=1}^K p_k=1$.\footnote{Note that we consider $\sigma_k^2$'s that are sufficiently small such that the probability for $\theta$ to exceed the $[-\frac{\pi}{2}, \frac{\pi}{2})$ region is negligible.} Under this model, the PDF of $\theta$ is given by
\begin{align}\label{angle}
	p_\Theta(\theta)=\sum_{k=1}^{K}p_k\frac{1}{\sqrt{2\pi}\sigma_k} e^{-\frac{(\theta-\theta_k)^2}{2\sigma_k^2}}.
\end{align}
Note that the considered Gaussian mixture model can characterize a wide range of practical scenarios by choosing different parameters. For example, when $K$ is sufficiently large, the PDF will tend to be uniform; while when $K=1$, the PDF will reduce to the Gaussian PDF.

Let $f_{k}(\theta)=\frac{1}{\sqrt{2\pi}\sigma_k} e^{-\frac{(\theta-\theta_k)^2}{2\sigma_k^2}}$ denote each $k$-th Gaussian PDF in the Gaussian mixture model. Then, $[\bm{F}_p]_{1,1}$ can be expressed as
\begin{align}\label{Fp11}
	&[\bm{F}_p]_{1,1}=\int  \left(\frac{\partial \ln(p_\Theta(\theta))}{\partial \theta}\right)^2\!\!p_\Theta(\theta)d\theta  \!\! \\
	&=\!\!\sum\limits_{k=1}^{K}\!\frac{p_k}{\sigma_k^2} \! -\!\!\!\!\underbrace{  \mathlarger{\int}  \frac{\sum\limits_{k_1=1}^{K}\!\sum\limits_{k_2=1 }^{K}\!p_{k_1}\!p_{k_2}\!f_{k_1}\!(\theta)\!f_{k_2}\!(\theta)\!\!\left(\!\!\frac{\theta\!-\!\theta_{k_1}}{\sigma_{k_1}^2}\!\!-\!\!\frac{\theta\!-\!\theta_{k_2}}{\sigma_{k_2}^2} \!\!\right)^{\!\!2}  }{ 2\sum\limits_{k=1}^{K}p_kf_k(\theta)}   d\theta}_\rho\!.  \nonumber
\end{align}
Note that $[\bm{F}_p]_{1,1}=\sum_{k=1}^K \frac{p_k}{\sigma_k^2}-\rho\geq 0$ holds according to (\ref{Fp11}).

Therefore, the overall FIM for $\boldsymbol{\zeta}$ is given by
\begin{align}
	\boldsymbol{F} =\bm{F}_o +\bm{F}_p= \left[
	\begin{array}{cc}
		J_{\theta\theta}+ \sum\limits_{k=1}^{K}\frac{p_k}{\sigma_k^2}-\rho            & \bm{J}_{\theta\alpha}     \\
		\bm{J}_{\theta\alpha}^H & \bm{J}_{\alpha\alpha}
	\end{array}
	\right].
\end{align}
Then, we have
\begin{align}
	\bm{F}^{-1}
	= \left[
	\begin{array}{ll}
		S^{-1} & \bm{C}       \\
		\bm{C}^H           & \bm{E}
	\end{array}
	\right],
\end{align}
where $S \in \mathbb{C}$, $\bm{C} \in \mathbb{C}^{1\times2}$, and $\bm{E} \in \mathbb{C}^{2\times2}$. Particularly, $S$ is the Schur complement of block $ \bm{J}_{\alpha\alpha}$, which is given by
\begin{align}\label{Fx}
	S \overset{\Delta}{=}
	J_{\theta\theta}+ \sum_{k=1}^{K}\frac{p_k}{\sigma_k^2}-\rho -  \bm{J}_{\theta\alpha} \bm{J}_{\alpha\alpha}^{-1} \bm{J}^H_{\theta\alpha} .
\end{align}

The PCRB for estimating the desired sensing parameter $\theta$ is only dependent on $S$, which is expressed as
\begin{align}\label{PCRB}
	\!&\mathrm{PCRB}_{\theta}=[\bm{F}^{-1}]_{1,1}=S^{-1} \nonumber\\
	\!&= \frac{1}{\sum\limits_{k=1}^{K}\frac{p_k}{\sigma_k^2}\!\!-\!\!\rho \!+\!\!\frac{2|\alpha|^2L}{\sigma_s^2}\!\left(\! \mathrm{tr}\left( (\bm{A}_1\!+\!\bm{A}_2)\bm{W}\right) \!-\!\frac{ \left| \mathrm{tr}(\bm{A}_3\bm{W})\right|^2}{ \mathrm{tr}(\bm{A}_4\bm{W})}\!\right)}.
\end{align}
Notice that $\mathrm{PCRB}_{\theta}$ is determined by the transmit covariance matrix $\bm{W}$, whose optimization will be studied in Section \ref{sec_pro}.

\subsection{Tractable Bound of PCRB}
Note that the exact PCRB in (\ref{PCRB}) has a complicated expression, which is difficult to analyze and draw insights from; moreover, the matrices $\bm{A}_1$, $\bm{A}_2$, $\bm{A}_3$, and $\bm{A}_4$ in the PCRB expression involve complicated integrals, obtaining which for the optimization of $\bm{W}$ requires high complexity. Motivated by the above, we propose an upper bound of the exact PCRB $\mathrm{PCRB}_{\theta}$, whose tightness will be verified numerically in Section \ref{sec_num}.

\begin{proposition}\label{prop_bound}
	$\mathrm{PCRB}_{\theta}$ is upper bounded as
	\begin{align}\label{bound}
		\mathrm{PCRB}_{\theta}\!\leq\! \mathrm{PCRB}_\theta^U \! \overset{\Delta}{=} \frac{1}{\sum\limits_{k=1}^{K}\frac{p_k}{\sigma_k^2}\!-\!\rho \!+\!\!\frac{2|\alpha|^2L}{\sigma_s^2} \mathrm{tr}(\bm{A}_1\bm{W})}.
	\end{align}
\end{proposition}
\begin{IEEEproof}
	Please refer to Appendix \ref{proof_bound}.
\end{IEEEproof}

Notice that the PCRB upper bound in (\ref{bound}) is in a much simpler form compared to the exact PCRB in (\ref{PCRB}). In the following, we will leverage this upper bound for discussing the effect of exploiting prior information in the estimation of $\theta$, and for optimizing the transmit covariance matrix $\bm{W}$.
\subsection{Effect of Exploiting Prior Information}
In this subsection, we aim to investigate the effect of exploiting prior distribution information on the sensing performance. Specifically, when the prior distribution information of $\theta$ is unknown, CRB can be adopted to characterize a lower bound of the estimation MSE corresponding to each realization of $\theta$. Given a realization of $\theta$, the FIM for estimating $\boldsymbol{\zeta}$ is given by $\tilde{\bm{F}} =\mathbb{E}_{\bm{Y} }\left[\frac{\partial \ln(f(\bm{Y}|\boldsymbol{\zeta}))}{\partial \boldsymbol{\zeta} }\left(\frac{\partial \ln(f(\bm{Y}|\boldsymbol{\zeta}))}{\partial \boldsymbol{\zeta} }\right)^H\right]$. The CRB corresponding to the given realization of $\theta$ is thus given by
\begin{align}\label{CRB}
	\mathrm{CRB}_{\theta}(\theta)\!= [\tilde{\bm{F}}^{-1}]_{1,1}=\!\frac{\frac{\sigma_s^2}{2|\alpha|^2L }}{ \|\dot{\bm{b}}(\theta)\|^2 \mathrm{tr}\left(\bm{a}(\theta)\bm{a}^H(\theta)\bm{W} \right)}.
\end{align}
Moreover, by taking the expectation of $\mathrm{CRB}_{\theta}(\theta)$ over the random angle realizations, the average (expected) CRB is given by
\begin{align}
	\mathrm{CRB}_{\theta} =\mathbb{E}_{\theta}[\mathrm{CRB}_{\theta}(\theta)]=\int \mathrm{CRB}_{\theta}(\theta)   p_\Theta(\theta)d\theta,
\end{align}
which can be viewed as a lower bound of the long-term MSE performance without exploiting prior information. Note that based on Jensen's inequality and $\sum_{k=1}^K \frac{p_k}{\sigma_k^2}-\rho\geq 0$, we have
\begin{align}
	\mathrm{CRB}_{\theta}=&\mathbb{E}_{\theta}\left[\frac{\frac{\sigma_s^2}{2|\alpha|^2L}}{\|\dot{\bm{b}}(\theta)\|^2\mathrm{tr}\left(\bm{a}(\theta)\bm{a}^H(\theta)\bm{W}\right)}\right]\nonumber\\
	\geq&\frac{1}{\mathbb{E}_{\theta}\left[\frac{\|\dot{\bm{b}}(\theta)\|^2\mathrm{tr}\left(\bm{a}(\theta)\bm{a}^H(\theta)\bm{W}\right)}{\frac{\sigma_s^2}{2|\alpha|^2L}}\right]}=\frac{\frac{\sigma_s^2}{2|\alpha|^2L}}{ \mathrm{tr}\left(\bm{A}_1\bm{W}\right)}\nonumber\\
	\geq&\mathrm{PCRB}_\theta^U\geq \mathrm{PCRB}_{\theta}.
\end{align}

The above result indicates that exploiting the prior distribution information can achieve a \emph{decreased} lower bound on the estimation MSE. Since the CRB/PCRB is generally tight in the moderate-to-high SNR regime, this further implies that the estimation performance can be improved via the exploitation of prior information. Moreover, according to the properties of Jensen's inequality \cite{arxiv_gap}, the gap between $\mathrm{CRB}_{\theta}$ and $\mathrm{PCRB}_{\theta}$ generally increases as the variance of $\theta$ increases. For example, if $\sigma_k^2$'s are fixed, for scenarios where the locations $\{\theta_1,...,\theta_K\}$ are more dispersed, the performance gain via exploiting prior distribution information will be more significant.

\section{Problem Formulation}\label{sec_pro}
In this section, we formulate the problem of optimizing the transmit covariance matrix $\bm{W}$ towards the optimal trade-off between communication and sensing. Specifically, we aim to minimize the PCRB of sensing the azimuth angle in (\ref{PCRB}), subject to a minimum communication rate threshold denoted by $\bar{R}$ bps/Hz. The optimization problem is formulated as
\begin{align} \label{P1}
	\!\!\!\mbox{(P1)}\!\!\!\!\quad  \mathop{\mathrm{min}}_{\bm{W} } \quad  & \mathrm{PCRB}_{\theta}
	\\
	\mathrm{s.t.} \quad & \log_2\left|\bm{I}_{N_u}+\frac{\bm{H}\bm{W}\bm{H}^H}{\sigma_c^2}\right| \geq \bar{R} \label{rate_constraint}\\
	& \mathrm{tr}\left( \bm{W}\right)\leq P  \label{power_constraint}\\
	&  \bm{W} \succeq \bm{0} \label{W_constraint}.
\end{align}

It is worth noting that (P1) is a non-convex optimization problem since the PCRB can be shown to be a non-convex function of $\bm{W}$ which involves a complicated fractional structure. Particularly, the optimal solution to (P1) even without the communication rate constraint, i.e., the \emph{sensing-optimal transmit covariance matrix} when the prior PDF is available for exploitation, still remains unknown. In the following, we will address the above challenges and derive the optimal solution to (P1); then, we will also propose a low-complexity suboptimal solution by leveraging the PCRB upper bound in (\ref{bound}).

\section{Proposed Solutions To Problem (P1)}\label{sec_sol}
In this section, we will first check the feasibility of (P1), and then present the optimal solution and a suboptimal solution.
\subsection{Feasibility of Problem (P1)}
Before solving (P1), we first check its feasibility.  Note that (P1) is feasible if and only if there exists a $\bm{W}$ that satisfies constraints (\ref{rate_constraint})-(\ref{W_constraint}),  which implies that the capacity of the MIMO channel from the BS to the user denoted by $R_{\max}$ is no smaller than the rate target $\bar{R}$. Specifically, $R_{\max}$ can be obtained by solving the following optimization problem:
\begin{align} \label{check}
	\mbox{(P1-F)}\quad  \mathop{\mathrm{max}}_{\bm{W} \succeq \bm{0} } \quad  & \log_2\left|\bm{I}_{N_u}+ \frac{\bm{H}\bm{W}\bm{H}^H}{\sigma_c^2}\right|\\
	\mathrm{s.t.} \quad  & \mathrm{tr}\left( \bm{W}\right)\leq P.
\end{align}
Note that the optimal solution to (P1-F) can be obtained according to the (reduced) singular value decomposition (SVD) of the MIMO channel matrix $\bm{H}$ \cite{Elements}. Denote the SVD of $\bm{H}$ as $\bm{H}=\bm{U}\bm{\Gamma}^{\frac{1}{2}}\bm{V}^H$, where $\bm{\Gamma}=\mathrm{diag}\{h_1,...,h_{T}\}$ with $T=\mathrm{rank}(\bm{H})$ and $h_1\geq h_2\geq...\geq h_{T}$;  $\bm{U} \in \mathbb{C}^{N_u\times T}$ and  $\bm{V}^H \in \mathbb{C}^{T\times N_t}$ are unitary matrices with $\bm{UU}^H= \bm{I}_{N_u} $ and $\bm{VV}^H= \bm{I}_{N_t}$. The optimal solution to (P1-F), i.e., the \emph{communication-optimal transmit covariance matrix}, can be expressed as $\bm{W}_c^\star=\bm{V}\bm{\Lambda}\bm{V}^H$ where $\bm{\Lambda}=\mathrm{diag}\{v_1,...,v_{T}\}$, with $v_i=\left(\nu-\sigma_c^2/h_i\right)^+,i=1,...,T$ denoting the water-filling based power allocation. The corresponding MIMO channel capacity is thus given by
\begin{align}
	R_{\mathrm{max}}=\sum_{i=1}^{T}\log_2\left(1+\frac{v_ih_i}{\sigma_c^2}\right).
\end{align}

Therefore, Problem (P1) is feasible if and only if $\bar{R}\leq R_{\max}$. In the following, we will solve (P1) assuming that it has been verified to be feasible.

\subsection{Optimal Solution to Problem (P1)}\label{sec_solution}
Note that $\bm{W}$ only affects the denominator in $\mathrm{PCRB}_\theta$ shown in	(\ref{PCRB}), the minimization of which is equivalent to the maximization of $\mathrm{tr}( (\bm{A}_1+\bm{A}_2)\bm{W}) -\frac{ \left| \mathrm{tr}(\bm{A}_3\bm{W})\right|^2}{ \mathrm{tr}(\bm{A}_4\bm{W})}$. Hence, Problem (P1) is equivalent to Problem (P2), which is given by
\begin{align} \label{P2}
	\!\!\!\mbox{(P2)}\!\!\!\!\quad  \mathop{\mathrm{max}}_{\bm{W} } \quad \!\!\!& \mathrm{tr}( (\bm{A}_1+\bm{A}_2)\bm{W}) -\frac{ \left| \mathrm{tr}(\bm{A}_3\bm{W})\right|^2}{ \mathrm{tr}(\bm{A}_4\bm{W})}
	\\
	\mathrm{s.t.} \quad  & \log_2\left|\bm{I}_{N_u}+ \frac{\bm{H}\bm{W}\bm{H}^H}{\sigma_c^2}\right| \geq \bar{R}\label{P2_C1}  \\
	& \mathrm{tr}\left( \bm{W}\right)\leq P  \\
	&  \bm{W} \succeq \bm{0}.\label{P2_C3}
\end{align}

By introducing an auxiliary variable $t$, we first transform (P2) into an equivalent problem:
\begin{align}
	\mbox{(P2')}\  \mathop{\mathrm{max}}_{\bm{W} ,t} \  & t\\
	\mathrm{s.t.}\ & \mathrm{tr}( (\bm{A}_1\!\!+\!\!\bm{A}_2)\bm{W})\!\! -\!\!\frac{ \left| \mathrm{tr}(\bm{A}_3\bm{W})\right|^2}{ \mathrm{tr}(\bm{A}_4\bm{W})}-t\geq0 \label{auxiliary_constraint}\\
	&   (\ref{P2_C1})-(\ref{P2_C3}). \nonumber
\end{align}
Then, we apply the Schur complement technique \cite{schur} to transform (P2') to an equivalent convex problem. Specifically, we express $\bm{W}= \sum_{i=1}^{N_t}u_i\bm{w}_i\bm{w}_i^H$ where $u_i\geq0,\ i=1,...,N_t$  and $\bm{w}_i \in \mathbb{C}^{N_t\times 1},\ i=1,...,N_t$ denote the eigenvalues and eigenvectors in $\bm{W}$, respectively. Based on this, we have
\begin{align}\label{A4}
	\mathrm{tr}\left(\bm{A}_4\bm{W}\right) &=N_r\int \sum_{i=1}^{N_t}u_i\mathrm{tr}\left(\bm{a}(\theta)\bm{a}^H(\theta)\bm{w}_i\bm{w}_i^H\right) p_\Theta(\theta)d\theta \nonumber\\
	&=N_r\int \sum_{i=1}^{N_t}u_i|\bm{a}^H(\theta)\bm{w}_i|^2  p_\Theta(\theta)d\theta > 0.
\end{align}
Note that $\mathrm{tr}( (\bm{A}_1+\bm{A}_2)\bm{W}) -\frac{ \left| \mathrm{tr}(\bm{A}_3\bm{W})\right|^2}{ \mathrm{tr}(\bm{A}_4\bm{W})} \!-\!t$ is the Schur complement of $ \mathrm{tr}\left(\bm{A}_4\bm{W}\right)$ in matrix $\bm{B}(t,\bm{W})=\left[\!\!\!
\begin{array}{ll}
	\mathrm{tr}\left( (\bm{A}_1+\bm{A}_2) \bm{W}\right) -t           & \!\! \mathrm{tr}\left(\bm{A}_3\bm{W}\right)    \\
	\mathrm{tr}\left(\bm{A}_3^H\bm{W}\right) & \!\! \mathrm{tr}\left(\bm{A}_4\bm{W}\right)
\end{array}
\!\!\!\right]$. Based on the Schur complement condition \cite{schur}, $\mathrm{tr}( (\bm{A}_1+\bm{A}_2)\bm{W}) -\frac{ \left| \mathrm{tr}(\bm{A}_3\bm{W})\right|^2}{ \mathrm{tr}(\bm{A}_4\bm{W})}-t\geq0$ with $\mathrm{tr}\left(\bm{A}_4\bm{W}\right)>0$ is equivalent to  $\bm{B}(t,\bm{W})\succeq \bm{0}$. Therefore, (P2') and consequently (P2) and (P1) are equivalent to the following problem:
\begin{align} \label{P3}
	\!\!\!\!(\mbox{P3})\!\!\!\!\quad  \mathop{\mathrm{max}}_{\bm{W} ,t} \quad  & t
	\\
	\mathrm{s.t.}  \quad
	& \left[\begin{array}{ll}
		\mathrm{tr}\left((\bm{A}_1\!+\!\bm{A}_2) \bm{W}\right)\!-\!t  &\!\! \mathrm{tr}\left(\bm{A}_3\bm{W}\right) \\
		\mathrm{tr}\left(\bm{A}_3^H\bm{W}\right)                      &\!\! \mathrm{tr}\left(\bm{A}_4\bm{W}\right)
	\end{array} \right] \succeq \bm{0}  \label{P3_c1}\\
	& \log_2\left|\bm{I}_{N_u}+ \frac{\bm{H}\bm{W}\bm{H}^H}{\sigma_c^2}\right|\geq \bar{R}\label{P3_c2} \\
	& \mathrm{tr}\left( \bm{W}\right)\leq P\label{P3_c3}  \\
	& \bm{W} \succeq \bm{0}.\label{P3_c4}
\end{align}

Note that Problem (P3) is a convex optimization  problem, for which the optimal solution can be obtained via the interior-point method \cite{CVX} or existing software, e.g., CVX \cite{cvxtool}. The optimal solution of $\bm{W}$ to Problem (P1) can be consequently obtained as the optimal solution of $\bm{W}$ to Problem (P3).

\begin{remark}[Optimal Solution to (P1) with $N_t=1$]
	Consider the case of $N_t=1$, where $\bm{W}$ reduces to a real value $W \in \mathbb{R}$ with $0\leq W\leq P$ that represents the transmit power. In this case, $\bm{A}_1\!=\!\int \|\dot{\bm{b}}(\theta)\|^2 p_\Theta(\theta)d\theta $, $\bm{A}_2\!=\!0$, $\bm{A}_3\!=\!0$, and $\bm{A}_4=N_r$. Thus, $\mathrm{PCRB}_{\theta}$ can be rewritten as $\mathrm{PCRB}_{\theta} = 1\Big/\left(\sum\limits_{k=1}^{K}\frac{p_k}{\sigma_k^2}\!-\!\rho \!+\!\!\frac{2|\alpha|^2L}{\sigma_s^2}W\int \|\dot{\bm{b}}(\theta)\|^2 p_\Theta(\theta)d\theta \right)$, which is a \emph{monotonically decreasing} function of $W$. The BS-user channel reduces to $\bm{H} =\bm{h}_u\in \mathbb{C}^{N_u\times 1}$, which yields a communication rate of $ R =  \log_2\left|\bm{I}_{N_u}+ \frac{W\bm{h}_u \bm{h}_u^H}{\sigma_c^2}\right|=\log_2\left(1+ \frac{W\|\bm{h}_u\|^2}{\sigma_c^2}\right)$, which is a \emph{monotonically increasing} function with respect to $W$. Thus, the optimal solution to (P3) and (P1) is $W^\star=P$.
\end{remark}
\subsection{Properties of the Optimal Solution}
To draw useful insights on the optimal dual-function transmit covariance matrix design in ISAC exploiting prior information, we explore the properties of the optimal solution via the Lagrange duality method.

First, we introduce dual variables $\bm{Z}_B=[z_1,z_2;z_2^*,z_3]\succeq \bm{0}$, $\mu_R\geq0$, $\mu_P\geq0$, and $\bm{Z}_W \succeq \bm{0}$, which are associated with constraints (\ref{P3_c1})-(\ref{P3_c4}), respectively. The Lagrangian of (P3) is thus given by
\begin{align}\label{P3_LA}
	&\mathcal{L}(t,\bm{W},\mu_P,\mu_R,\bm{Z}_W,\bm{Z}_B)\!=\!t\!+\!\mathrm{tr}\left(\bm{Z}_B\bm{B}(t,\bm{W})\right)\!+\!\mathrm{tr}\left(\bm{Z}_W\bm{W}\right)
	\nonumber \\
	&\!-\!\mu_P( \mathrm{tr}( \bm{W})\!-\!P)
	\!+\!\mu_R\left(  \log_2\left|\bm{I}_{N_u}\!+\! \frac{\bm{H}\bm{W}\bm{H}^H}{\sigma_c^2}\right|\! - \!\bar{R} \right).\!\!
\end{align}
Specifically, $\mathrm{tr}\left(\bm{Z}_B\bm{B}(t,\bm{W})\right)$ can be rewritten as
\begin{align}
	\mathrm{tr}\left(\bm{Z}_B\bm{B}(t,\bm{W})\right)=-z_1t+\mathrm{tr}\left(\bm{D}\bm{W}   \right),
\end{align}
where $\bm{D}=z_1\left(\bm{A}_1+\bm{A}_2\right)+z_2\bm{A}^H_3+z_2^*\bm{A}_3+z_3\bm{A}_4$. The Karush-Kuhn-Tucker (KKT) optimality conditions include (\ref{P3_c1})-(\ref{P3_c4}) and:
\begin{align}
	\mathrm{tr}\left(\bm{Z}_B\bm{B}(t,\bm{W})\right)=0\label{KKT1}\\
	\mu_R\left(\log_2\left|\bm{I}_{N_u}+\frac{\bm{H}\bm{W}\bm{H}^H}{\sigma_c^2}\right|-\bar{R}\right)=0\\
	\mu_P(\mathrm{tr}(\bm{W})-P)=0\\
	\mathrm{tr}(\bm{Z}_W \bm{W})=0\\
	\frac{\partial \mathcal{L}(t,\bm{W},\mu_P,\mu_R,\bm{Z}_W,\bm{Z}_B)}{\partial t}=1-z_1=0.\label{derivative_1}
\end{align}
Let $t^\star$, $\bm{W}^\star$, $\bm{Z}_B^\star=[z_1^\star,z_2^\star;z_2^{\star^*},z_3^\star]$, $\mu_R^\star$, $\mu_P^\star$, and $\bm{Z}_W^\star$ denote the optimal primal and dual solutions. Based on (\ref{derivative_1}), we have $z^\star_1=1$, which implies that $\bm{Z}^\star_B \neq \bm{0}$. Based on (\ref{KKT1}), we have $z^\star_1z^\star_3-|z^\star_2|^2=0$. Thus, we have $\bm{Z}_B^\star=[1,z_2^\star;z_2^{\star^*},|z_2^\star|^2]$, and $\bm{D}^\star$ can be expressed as
\begin{align}\label{matrixD}
	&\bm{D}^\star=   \bm{A}_1+\bm{A}_2 +z^\star_2\bm{A}^H_3+z_2^{\star^*}\bm{A}_3+|z^\star_2|^2\bm{A}_4 \\
	&=\!\!\int\!\! \left[\bm{a}(\theta),   \dot{\bm{a}}(\theta)\right]\!\!
	\left[ \begin{array}{ll}
		\!\!\|\dot{\bm{b}}(\theta)\|^2+ |z^\star_2|^2N_r &\!\!\!\! z^\star_2N_r  \\
		\!\!z^{\star^*}_2 N_r &\!\!\!\! N_r
	\end{array}\!\!\right]\!\!\!
	\left[ \begin{array}{l }
		\!\!\bm{a}^H(\theta)    \\
		\!\!\dot{\bm{a}}^H(\theta)
	\end{array}\!\!\right]\!\! p_\Theta(\theta)d\theta,\nonumber
\end{align}
which satisfies $\bm{D}^\star \succeq \bm{0}$.  We express the eigenvalue decomposition (EVD) of  $\bm{D}^\star$ as $\bm{D}^\star=\bm{Q}_D\bm{\Lambda}_D\bm{Q}_D^H$, where $\bm{\Lambda}_D=\mathrm{diag}\{d_1,...,d_{N_t}\}$ with $d_1\geq d_2\geq ...\geq d_{N_t}\geq 0$; $\bm{Q}_D=[\bm{q}_1,...,\bm{q}_{N_t}]$ is a unitary matrix with $\bm{Q}_D\bm{Q}_D^H=\bm{Q}_D^H\bm{Q}_D=\bm{I}_{N_t}$. Based on this, we analyze the optimal $\bm{W}^\star$ considering two cases where the communication rate constraint in (\ref{P3_c2}) is inactive or active, respectively.
\subsubsection{Case I} \emph{$\mu^\star_R=0$, i.e., the communication rate constraint is inactive.} In this case, we have the following proposition.
\begin{proposition}
	When $\mu^\star_R=0$, there exists a rank-one optimal solution to (P3) and (P1) given by $\bm{W}^\star=P\bm{q}_1\bm{q}^H_1$.
\end{proposition}
\begin{IEEEproof}
	In this case, the optimal solution of $\bm{W}$ can be obtained by solving the following problem:
	\begin{align}
		\mbox{(P3-I)} \quad \mathop{\mathrm{max}}_{\bm{W}\succeq \bm{0} } \quad & \mathrm{tr}\left(\bm{D}^\star \bm{W} \right)\\
		\mathrm{s.t.} \quad & \mathrm{tr}\left( \bm{W}\right)\leq P .
	\end{align}
	Since (P3-I) is a semi-definite program (SDP) with one linear constraint, there exists a rank-one optimal solution to (P3-I) and consequently (P3) \cite{Palomar}. Based on Proposition 2 in \cite{ISIT}, we have $\bm{W}^\star=P\bm{q}_1\bm{q}^H_1$.
\end{IEEEproof}

Let $\bm{W}^\star_s$ denote the optimal solution to (P3) and consequently (P1) without the communication constraint (i.e., with $\bar{R}=0$), which is the \emph{sensing-optimal transmit covariance matrix} that minimizes the PCRB and can be obtained by solving (P3-I). Let $R_s=\log_2\left|\bm{I}_{N_u}+\frac{\bm{H}\bm{W}_s^\star\bm{H}^H}{\sigma_c^2}\right|$ denote the communication rate with $\bm{W}^\star_s$. Note that when $\bar{R}\leq R_s$, the communication rate constraint is automatically inactive, thus we have the following corollary.
\begin{corollary}
	When $\bar{R}\leq R_s$, an optimal transmit covariance matrix solution to (P1) is given by $\bm{W}^\star=P\bm{q}_1\bm{q}^H_1$.
\end{corollary}

The above results imply that when the communication rate constraint is inactive (i.e., when the rate threshold $\bar{R}$ is low), the optimal transmit covariance matrix can have a \emph{rank-one structure}, although the transmit signals need to cater to a continuous range of possible target angles for sensing.

\subsubsection{Case II}\emph{$\mu^\star_R>0$, i.e., the communication constraint is active.} In this case, we have the following proposition.
\begin{proposition}\label{prop_rank}
	When $\mu_R^\star>0$, the optimal solution to (P3) and (P1) satisfies $\mathrm{rank}(\bm{W}^\star)\leq \mathrm{rank}(\bm{H})$.
\end{proposition}
\begin{IEEEproof}
	Please refer to Appendix \ref{proof_rank}.
\end{IEEEproof}

The results show that although the transmit signals need to cater to both the multi-antenna communication user and a sensing target under a continuous angle PDF with an infinitely large number of possible locations, the rank of the optimal transmit covariance matrix is still limited by the rank of the MIMO communication channel.

\subsection{Suboptimal Solution to Problem (P1)}
To further reduce the complexity of the optimal solution and to draw more useful insights, we propose to minimize the PCRB upper bound $\mathrm{PCRB}_\theta^U$  derived in (\ref{bound}) for finding an approximate solution to (P1). Since $\mathrm{PCRB}_\theta^U$ is a monotonically decreasing function with respect to $\mathrm{tr}\left(\bm{A}_1\bm{W}\right)$, minimizing $\mathrm{PCRB}_\theta^U$  is equivalent to maximizing $\mathrm{tr}\left(\bm{A}_1\bm{W}\right)$. Thus, the approximation of (P1) is given by
\begin{align} \label{P4}
	\!\!\!\!\mbox{(P4)}\!\!\!\!\quad  \mathop{\mathrm{max}}_{\bm{W}  }  \quad &  \mathrm{tr}\left(\bm{A}_1\bm{W}\right)
	\\
	\mathrm{s.t.}        \quad & \log_2\left|\bm{I}_{N_u}+ \frac{\bm{H}\bm{W}\bm{H}^H}{\sigma_c^2}\right|\geq \bar{R} \label{constraint} \\
	& \mathrm{tr}\left( \bm{W}\right)\leq P\label{P4c2}  \\
	&  \bm{W} \succeq \bm{0}.
\end{align}
(P4) is a convex problem. To reveal the optimal solution structure, we apply the Lagrange duality method to solve (P4) by discussing two cases where the communication rate constraint in (\ref{constraint}) is inactive or active, respectively.
\subsubsection{Case I} \emph{Inactive communication rate constraint.} In this case, (P4) reduces to an SDP. Based on Proposition 2 in \cite{ISIT}, the optimal solution to (P4) without constraint (\ref{constraint}) is $P\bm{s}_1\bm{s}_1^H$ with $\bm{s}_1 \in \mathbb{C}^{N_t\times 1}$ being the eigenvector corresponding to the largest eigenvalue of matrix $\bm{A}_1$, which is the optimal transmit covariance matrix for minimizing the PCRB upper bound. Denote $R_s^U=\log_2\left|\bm{I}_{N_u}+ \frac{P\bm{H}\bm{s}_1\bm{s}_1^H\bm{H}^H}{\sigma_c^2}\right|$ as its corresponding communication rate. If $R_s^U\geq\bar{R}$, the  constraint (\ref{constraint}) is inactive and the optimal solution to (P4) is given by $\bm{W}_U^\star=P\bm{s}_1\bm{s}_1^H$.
\subsubsection{Case II} \emph{Active communication rate constraint.} In this case, we have $R_s^U<\bar{R}$. Let $\beta>0$ and $\mu\geq0$ denote the dual variables associated with the constraints in (\ref{constraint}) and (\ref{P4c2}), respectively. The Lagrangian of (P4) is given by
\begin{align}
	\bar{\mathcal{L}}(\bm{W},\beta,\mu)\!= &\mathrm{tr}\left(\bm{A}_1\bm{W}\right)\!+\!\beta\!\left(\!\log_2\!\left|\bm{I}_{N_u}\!+\! \frac{\bm{H}\bm{W}\bm{H}^H}{\sigma_c^2}\right|\!\!-\!\!\bar{R}\right)\nonumber
	\\
	&-\mu\left( \mathrm{tr}\left(\bm{W}\right)-P \right), \; \bm{W}\succeq\bm{0}.
\end{align}
The Lagrangian dual function of (P4) is defined as $g(\beta,\mu)= \mathop{\mathrm{max}}\limits_{\bm{W} \succeq \bm{0}} \; \bar{\mathcal{L}}(\bm{W},\beta,\mu)$, and the dual problem of (P4) is defined as $\mathop{\mathrm{min}}\limits_{\beta >0,\mu\geq0}\; g(\beta,\mu)$. Note that given any $\bar{R}\in (R_s^U,R_{\mathrm{max}}]$, there exists unique optimal dual variables denoted by $\beta^\star$ and $\mu^\star$. The optimal solution to Problem (P4) can be then obtained via the following proposition.

\begin{proposition}\label{prop_solution}
	The optimal solution to (P4) with $R_s^U<\bar{R}$ is given by
	\begin{align}\label{solution}
		\bm{W}_U^\star= {\bm{Q}}^{-\frac{1}{2}}\tilde{\bm{V}}\tilde{\bm{\Lambda}}\tilde{\bm{ V}}^H{\bm{Q}}^{-\frac{1}{2}}.
	\end{align}
	Specifically, $\bm{Q}=\frac{\mu^\star}{\beta^\star}\bm{I}_{N_t}-\frac{1}{\beta^\star}\bm{A}_1$; $\bm{H}{\bm{Q}}^{-\frac{1}{2}}= \tilde{\bm{U}}\tilde{\bm{\Gamma}}^{\frac{1}{2}}\tilde{\bm{V}}^H$ denotes the (reduced) SVD of $\bm{H}{\bm{Q}}^{-\frac{1}{2}}$, where $\tilde{\bm{\Gamma}}=\mathrm{diag}\{\tilde{h}_1,...,\tilde{h}_{T}\}$, $\tilde{\bm{U}} \in \mathbb{C}^{N_u\times T}$ and $\tilde{\bm{V}}\in \mathbb{C}^{N_t\times T}$ are unitary matrices with $\tilde{\bm{U}}\tilde{\bm{U}}^H= \bm{I}_{N_u} $ and $\tilde{\bm{V}}\tilde{\bm{V}}^H= \bm{I}_{N_t}$; and $\tilde{\bm{ \Lambda}}=\mathrm{diag}\{\tilde{v}_1,...,\tilde{v}_{T}\}$ with $\tilde{v}_i=(1/\ln2-\sigma_c^2/\tilde{h}_i)^{+},\ i=1,2,...,T$.
\end{proposition}
\begin{IEEEproof}
	Please refer to Appendix \ref{proof_solution}.
\end{IEEEproof}

To summarize, the optimal solution to (P4) for achieving the best trade-off between the PCRB upper bound and the communication rate is given by
\begin{align} \label{sub}
	\boldsymbol{W}_U^\star \!\!=\!\! \left\{
	\begin{array}{ll}
		\!\!P\bm{s}_1\bm{s}_1^H,  &R_s^U\geq \bar{R},     \\
		\!\!{\bm{Q}}^{ -\frac{1}{2}}\tilde{\bm{V}}\tilde{\bm{\Lambda}}\tilde{\bm{ V}}^H{\bm{Q}}^{ -\frac{1}{2}},&R_s^U\!<\! \bar{R}\!\leq\! R_{\mathrm{max}}.
	\end{array}
	\right.
\end{align}

Note that when the communication constraint is inactive, the optimal transmit covariance matrix $\boldsymbol{W}_U^\star$ to (P4) has a \emph{rank-one structure}, and is solely dependent on the target angle's PDF $p_\Theta(\theta)$. On the other hand, when the communication constraint is active, $\bm{W}_U^\star$ is determined by both the PDF $p_\Theta(\theta)$ and the communication channel $\bm{H}$, with $\mathrm{rank}(\bm{W}_U^\star)\leq \mathrm{rank}(\tilde{\bm{\Lambda}})\leq \mathrm{rank}(\bm{H})$. Notice that the rank properties of the proposed suboptimal solution are consistent with those of the optimal solution.

\begin{remark}[Optimal Solution to (P4) with $N_u=1$]
	When $N_u=1$, the BS-user channel reduces to a MISO channel denoted by $\bm{H}=\bm{h}_t^H \in \mathbb{C}^{1\times N_t}$. We then have $R_s^U=\log_2\left(1+ \frac{P  |\bm{s}_1^H\bm{h}_t|^2}{\sigma_c^2}\right)$. The suboptimal solution to (P1) in (\ref{sub}) reduces to
	\begin{align}\label{MISO}
		\!\!\!	\bm{W}_U^\star \!\!=\!\! \left\{
		\begin{array}{ll}
			\!\!\!\!P\bm{s}_1\bm{s}_1^H,  \!\!\!\!\!&R_s^U\geq \bar{R},     \\
			\!\!\!\! {\bm{Q}}^{-\!1}\bm{h}_t\frac{(\|{\bm{Q}}^{ -\!\frac{1}{2}}\bm{h}_t\|^2 -\ln2)^+}{\ln2\left\|{\bm{Q}}^{  -\!\frac{1}{2}}\bm{h}_t\right\|^4}\bm{h}_t^H{\bm{Q}}^{-\!1}, \!\!\!\!\!&R_s^U\!\!<\!\! \bar{R}\leq R_{\mathrm{max}}.
		\end{array}
		\right.
	\end{align}
	Note that from (\ref{MISO}), the optimal transmit covariance matrix always has a \emph{rank-one structure} for the special case of $N_u=1$.
	
	On the other hand, in this case, we can obtain the optimal solution to (P4) in a direct manner without taking the inverse of $\bm{Q}$. Specifically, (P4) can be equivalently expressed as
	\begin{align}
		\mbox{(P4-MISO)}\quad   \mathop{\mathrm{max}}_{\bm{W}  } \quad & \mathrm{tr}\left(\bm{A}_1\bm{W}\right)
		\\
		\mathrm{s.t.}  \quad & \mathrm{tr}\left(\bm{h}_t\bm{h}_t^H\bm{W}\right)\geq \left(2^{\bar{R}}-1\right)\sigma_c^2\label{P4MISOrate}\\
		& \mathrm{tr}\left( \bm{W}\right) \leq P.
	\end{align}
	Let $\eta\geq 0$ denote the dual variable associated with (\ref{P4MISOrate}). The (partial) Lagrangian of (P4-MISO) is given by
	\begin{align}\label{Lagrangian}
		\tilde{\mathcal{L}}(\bm{W},\eta)\!=\!\mathrm{tr}\left(\left(\bm{A}_1\!+\!\eta\bm{h}_t\bm{h}_t^H\right)\bm{W}\right) \!\!-\!\!\eta\left(2^{\bar{R}}\!\!-\!1\right)\!\sigma_c^2,
	\end{align}
	with $\bm{W}\succeq\bm{0}$ and $\mathrm{tr}\left(\bm{W}\right) \leq P$. Given the optimal $\eta^\star$, the optimal solution to (P4-MISO) can be obtained by solving:
	\begin{align}
		\mbox{(P4-MISO')}\quad  \mathop{\mathrm{max}}_{\bm{W} \succeq \bm{0} }
		\quad & \mathrm{tr}\left(\left(\bm{A}_1+\eta^\star\bm{h}_t\bm{h}_t^H\right)\bm{W}\right)\\
		\mathrm{s.t.}\quad & \mathrm{tr}\left( \bm{W}\right) \leq P.
	\end{align}
	Based on Proposition 2 in \cite{ISIT}, the optimal solution to (P4-MISO') and (P4) with $N_u=1$ can be directly expressed as $\boldsymbol{W}_{U}^\star = P\tilde{\bm{s}}\tilde{\bm{s}}^H$, where $\tilde{\bm{s}}\in \mathbb{C}^{N_t\times 1}$ is the eigenvector corresponding to the largest eigenvalue of matrix $\bm{A}_1+\eta^\star\bm{h}_t\bm{h}_t^H$.
\end{remark}
\subsection{Complexity Analysis}
In this subsection, we analyze the complexities for obtaining the optimal solution and the suboptimal solution to (P1).

To obtain the optimal solution, we first need to calculate the matrices $\bm{A}_1$, $\bm{A}_2$, $\bm{A}_3$, and $\bm{A}_4$ via the integration operation. Let $\mathcal{O}(\varpi)$ denote the complexity of integrating a one-dimensional function over $[-\frac{\pi}{2},\frac{\pi}{2})$. The total complexity of obtaining these matrices is thus $\mathcal{O}(4N_t^2\varpi)$. Then, we analyze the complexity for using the interior-point method to solve (P3). Considering that there are three constraints, $N_t^2$ complex variables, and one real variable, the computational complexity is given as $\mathcal{O}( (3(2N_t^2+1)^2+(2N_t^2+1)^3)\sqrt{2N_t^2+1})$ \cite{CVX}, which can be rewritten as $\mathcal{O}(2^{3.5}N_t^7)$ by reserving the highest-order term. Thus, the computational complexity for finding the optimal solution is given by $\mathcal{O}(4N_t^2\varpi+ 2^{3.5}N_t^7)$.

To obtain the suboptimal solution via solving (P4) by Lagrange duality method, we only need to calculate one matrix $\bm{A}_1$, which requires a complexity of $\mathcal{O}(N_t^2\varpi)$. Let $N_{LD}$ represent the total number of iterations in the Lagrange duality method. In each iteration, we first obtain $\bm{Q}^{-\frac{1}{2}}$, for which the complexity is $\mathcal{O}(N_t^3)$ \cite{EVD}. Then, the solution is obtained based on the SVD of matrix $\bm{H}{\bm{Q}}^{-\frac{1}{2}}$, for which the complexity is $\mathcal{O}(N_uN_t\min(N_u,N_t) )$. Thus, the computational complexity for obtaining the suboptimal solution is $\mathcal{O}(N_t^2\varpi+N_{LD}(N_t^3+N_uN_t\min(N_u,N_t)))$, which is lower than that for finding the optimal solution.

\section{Numerical Results}\label{sec_num}
\begin{figure} [t]
	\centering
	\includegraphics[width=3.5in]{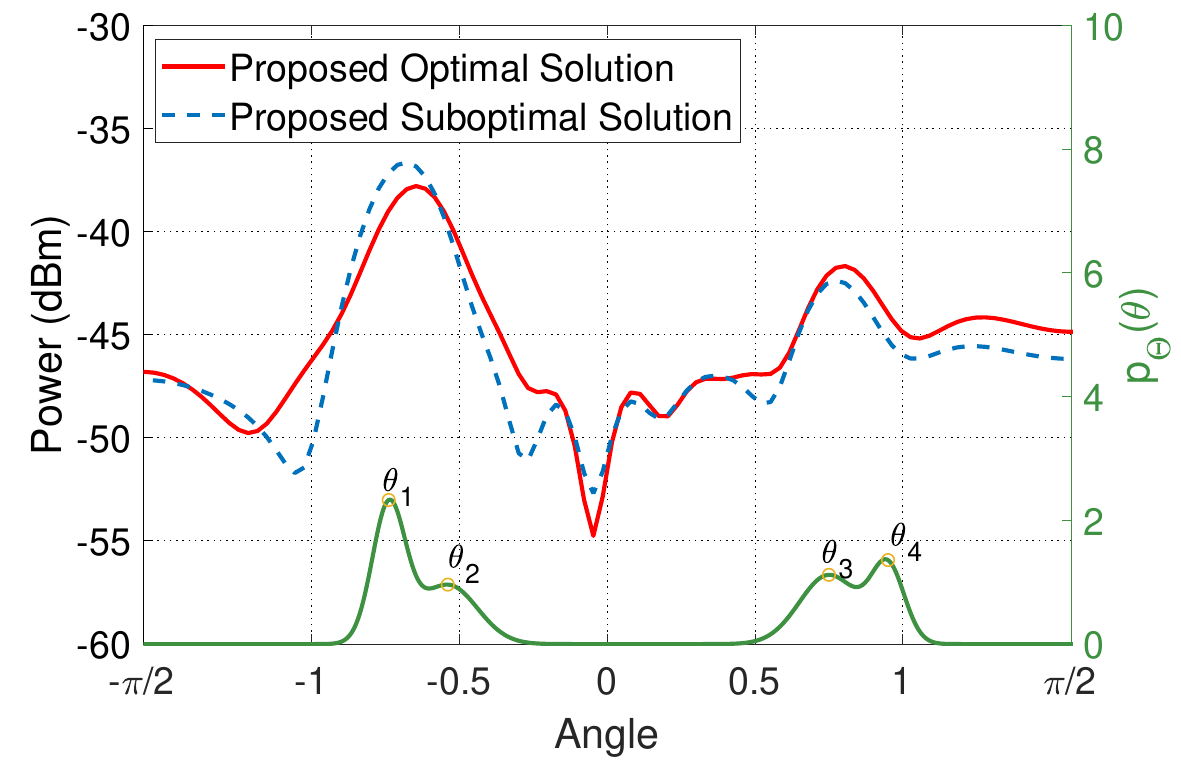}
	\vspace{-6mm}
	\caption{Illustration of the radiated power pattern and target angle PDF.}\label{Fig_Beampattern}
	\vspace{-6mm}	
\end{figure}

In this section, we provide numerical results to evaluate the performance of the proposed transmit covariance matrix designs for MIMO ISAC exploiting prior information. We set $N_r=12$, $N_u=8$, $L=25$, $P=30$ dBm, $\sigma_c^2=\sigma_s^2=-90$ dBm, $d=\frac{\lambda}{2}$, $h_B=10$ m. The known range of the target is set as $r=50$ m. We further set $N_t=10$, $\bar{R} =6.5$ bps/Hz, and $\frac{P|\alpha|^2L}{\sigma_s^2}=-5$ dB unless specified otherwise. In the target angle PDF $p_{\Theta}(\theta)$, we set $K=4$; $\theta_1=-0.74$, $\theta_2=-0.54 $, $\theta_3=0.75 $, $\theta_4=0.95 $; $\sigma_1^2 =10^{-2.5}$, $\sigma_2^2 =10^{-2}$, $\sigma_3^2 =10^{-2}$, $\sigma_4^2=10^{-2.5}$; and $p_1=0.31$, $p_2=0.24$, $p_3=0.28$, $p_4=0.17$. The corresponding PDF of the target's angle $\theta$ is illustrated in Fig. \ref{Fig_Beampattern}.

We consider a Rician fading model for the communication channel, which is given by $\bm{H}= \sqrt{\beta_c/(K_c+1)}(\sqrt{K_c}\bm{H}_{\text{LoS}}+\bm{H}_{\text{NLoS}})$. Specifically, $K_c=-8$ dB is the Rician factor and $\beta_c$ denotes the path loss given by $\beta_c= \beta_0/{r_U}^{\alpha_c} $, with $\beta_0=-30$ dB denoting the reference path loss at 1 m and $\alpha_c=3.5$ denoting the path loss exponent. The height of the user's antennas is set as $h_U\!=\!1$ m and $r_U\!=\!400$ m denotes the BS-user distance. The elevation angle of the user is thus given by $\phi_U=\arcsin\frac{h_U-h_B}{r_U}$. The LoS component is given by $\bm{H}_{\text{LoS}}=\bm{b}_U(\theta_U)\bm{a}(\theta_U)^H$ with $\theta_U$ being the azimuth angle of the user, $\bm{a}(\theta_U) \in \mathbb{C}^{N_t\times 1}$ and $\bm{b}_U(\theta_U)\in\mathbb{C}^{N_u\times 1}$ being the steering vectors of the BS transmit antenna array and the user receive antenna array, respectively. Specifically, by considering ULAs at two sides, we have $a_n(\theta_U)=e^{\frac{-j\pi d(N_t-2n+1)}{\lambda}\cos\phi_U\sin\theta_U},\forall n$ and ${b_U}_m(\theta_U)=e^{\frac{-j\pi d(N_u-2m+1)}{\lambda} \cos\phi_U\sin\theta_U},\forall m$, with $\cos\phi_U=\frac{\sqrt{r_U^2-(h_U-h_B)^2}}{r_U}$. The NLoS component $\bm{H}_{\text{NLoS}}$ is modeled by i.i.d. Rayleigh fading, with $[\bm{H}_{\text{NLoS}}]_{i,j} \sim \mathcal{CN}(0, 1),\ \forall i,j$. In the following, we set $\theta_U=0.36$.

\subsection{Comparison Between Optimal and Suboptimal Solutions}
First, we show in Fig. \ref{Fig_Beampattern} the radiated power pattern at distance $r$ over different angles with the proposed solutions. Note that the beampatterns achieved by both the optimal and suboptimal solutions are concentrated over angles with high probability densities, which demonstrates the effectiveness of both schemes in achieving probability-dependent power focusing for sensing, by judiciously designing the transmit covariance matrix based on the prior distribution information. Moreover, the two schemes achieve similar beampatterns, which validates the efficacy of the PCRB upper bound.

\begin{figure} [t]
	\centering
	\includegraphics[width=3in]{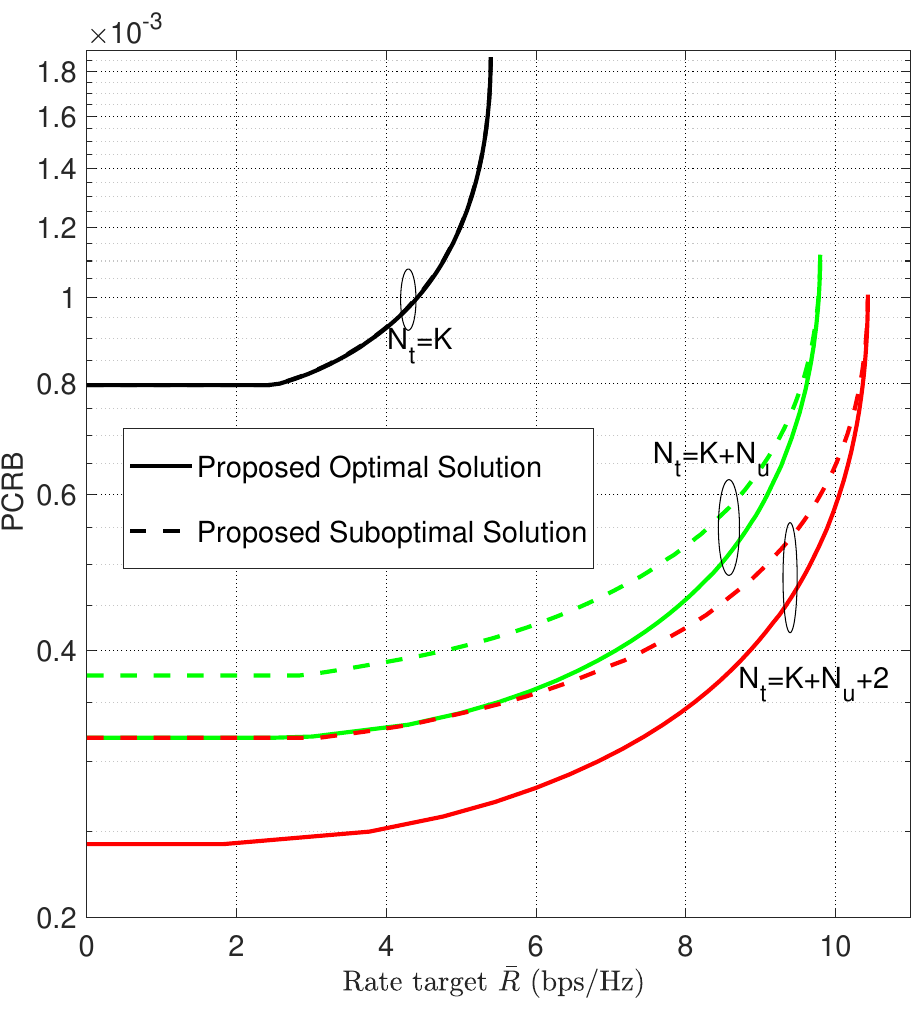}
	\vspace{-3mm}
	\caption{PCRB versus communication rate target under different numbers of transmit antennas.}\label{Fig_lowerbound2}
\end{figure}

\begin{figure} [t]
\centering
\includegraphics[width=3.5in]{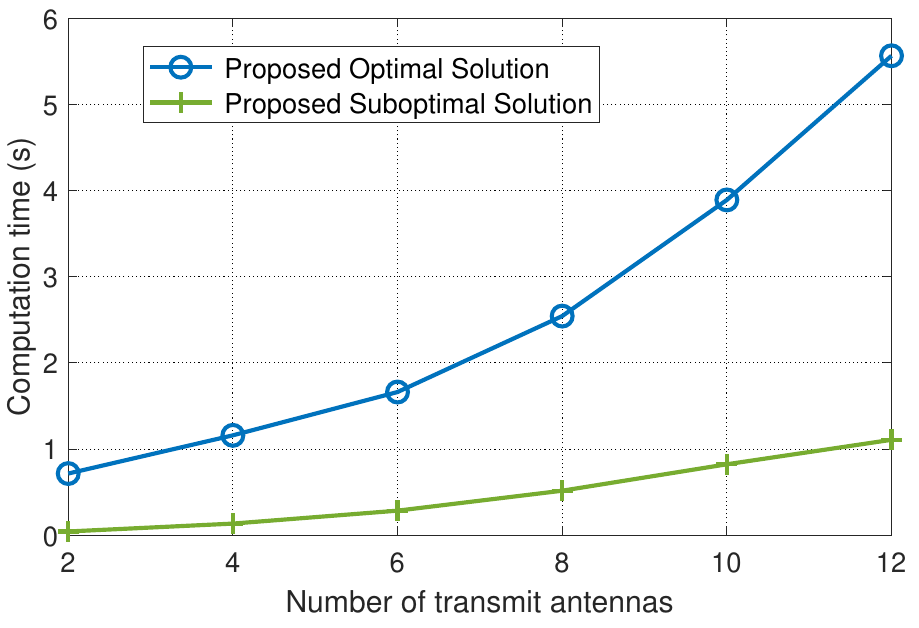}
\vspace{-6mm}
\caption{Computation time versus the number of transmit antennas.}\label{Fig_time}
\end{figure}

Then, Fig. \ref{Fig_lowerbound2} illustrates the PCRB performance achieved by the proposed solutions versus the communication rate target under different numbers of transmit antennas $N_t$. It is observed that there exists a rate-PCRB trade-off, and both the rate and PCRB can be improved by increasing $N_t$, due to the larger spatial degrees-of-freedom (DoFs) available at the transmitter. Moreover, the performance gap between the optimal and suboptimal solutions increases as $N_t$ increases. On the other hand, we show in Fig. \ref{Fig_time} the required computation time for obtaining the optimal and suboptimal solutions via MATLAB on a computer with an Intel Core i7 3.20-GHz CPU and 18 GB of memory. It is observed that the proposed suboptimal solution requires much less computation time than the optimal solution especially with large $N_t$, due to fewer integration operations needed and the derived semi-closed-form expression of the suboptimal solution. In practice, the optimal and suboptimal solutions can be chosen flexibly based on the performance and complexity requirements.


\begin{figure} [t]
	\centering
	\includegraphics[width=3in]{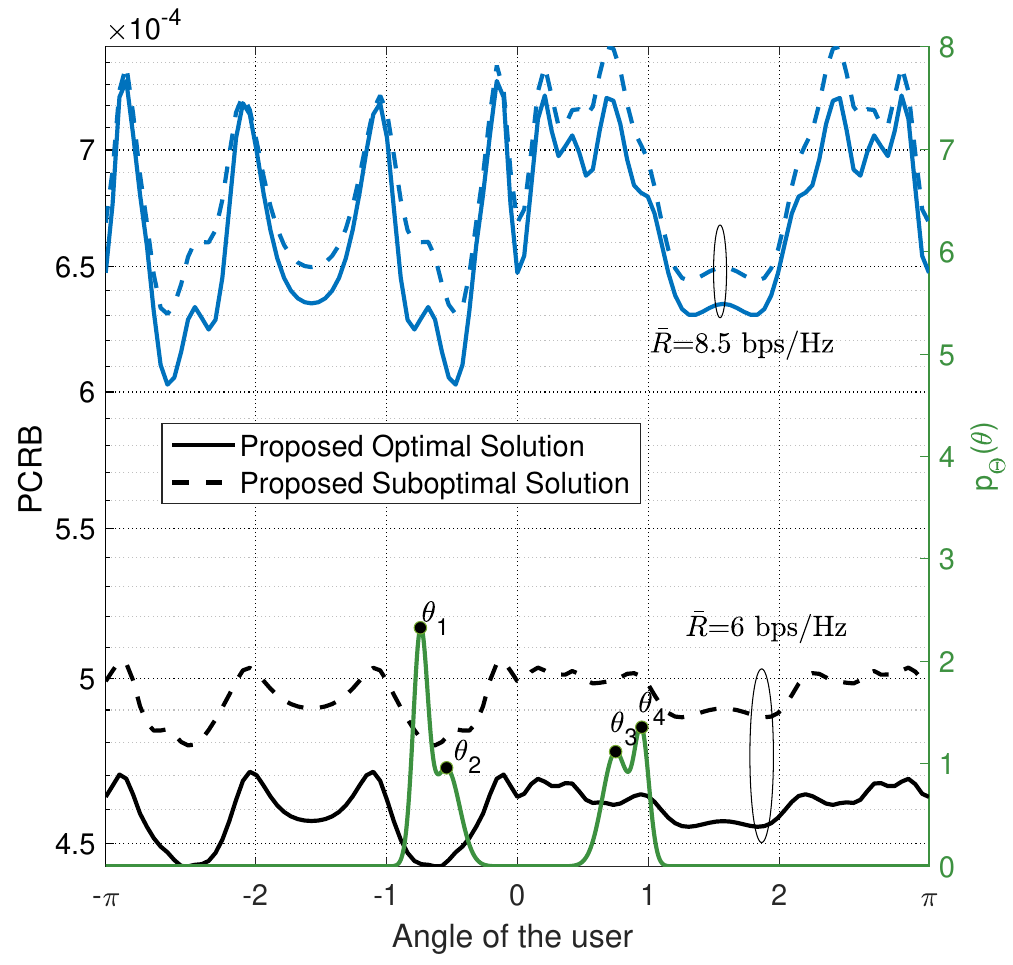}
	\vspace{-3mm}
	\caption{PCRB versus angle of the communication user.}\label{Fig_angle}
	\vspace{-3mm}	\end{figure}

Finally, we show in Fig. \ref{Fig_angle} the PCRB achieved by the proposed solutions versus the azimuth angle of the user. It is observed that when the user's azimuth angle $\theta_U$ approaches the target's possible angles with high probability densities (e.g., $\theta_1$ and $\theta_2$) or their symmetric angles,\footnote{Due to the symmetry of ULA, the steering vector of the BS antennas over an angle is the same as that over its symmetric angle.} the PCRB tends to be small even with a high rate target $\bar{R}$, since the signals focused around these angles can be more efficiently used for both sensing and communication. This suggests that the BS responsible for target sensing should be selected as the one whose associated communication users are located closely to the target's highly probable locations.

\subsection{Comparison with Genie-Aided Benchmark Schemes}

\begin{figure} [t]
	\centering
	\includegraphics[width=3.5in]{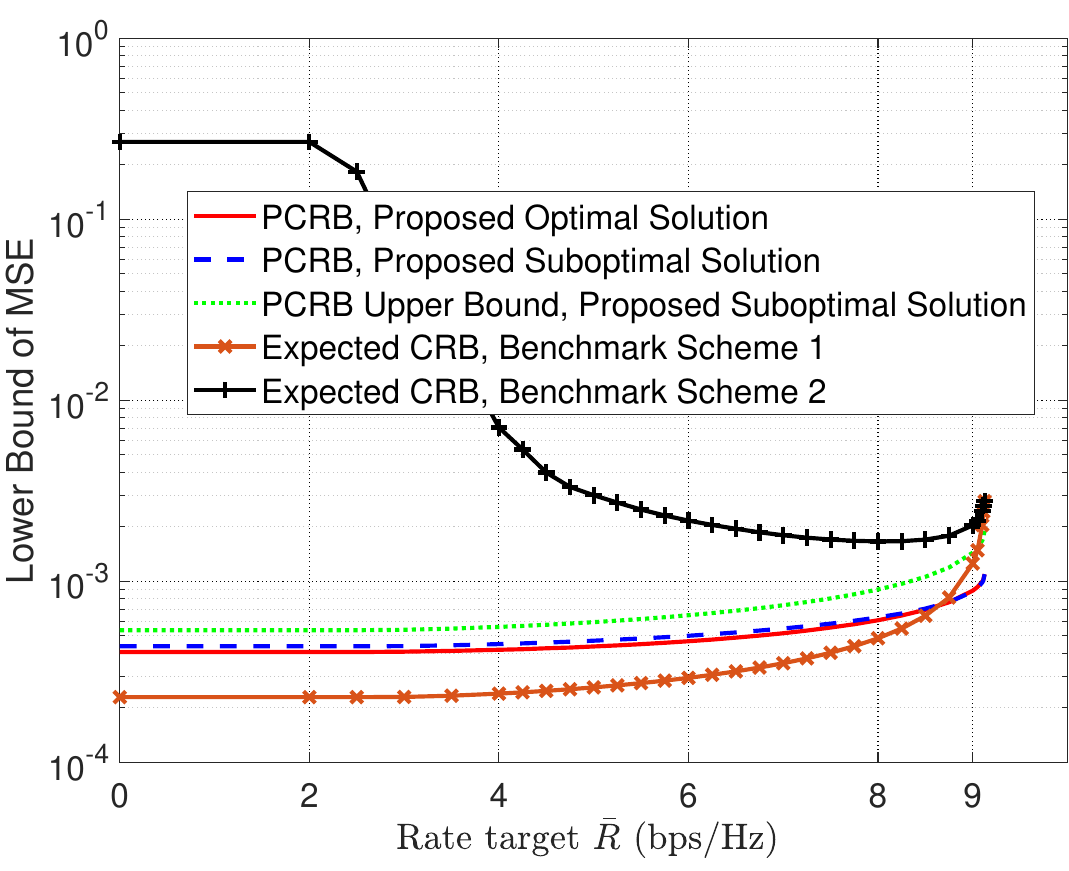}
	\vspace{-6mm}
	\caption{Lower bound of MSE (PCRB/expected CRB) versus the communication rate target with different transmit covariance matrix designs.}\label{Fig_lowerbound1}
		\end{figure}

Next, for comparison, we consider the following benchmark schemes for the transmit covariance matrix design. Specifically, the two schemes assume that the exact or inexact value of each realization of $\theta$ is known, which are genie-aided schemes.
\begin{itemize}
	\item {\bf{Benchmark Scheme 1: Exact target location based transmit covariance matrix design}.} In this scheme, the exact value of each realization of $\theta$ is known and denoted by $\theta_e$. Then, the CRB $\mathrm{CRB}_\theta(\theta_e)$ in (\ref{CRB}) is minimized under the constraints in (P1), which is a convex problem.
	\item {\bf{Benchmark Scheme 2: Inexact target location based transmit covariance matrix design}.} In this scheme, an inexact value of $\theta$ denoted by $\tilde{\theta}$ is known, which follows a distribution of $\tilde{\theta}\sim\mathcal{N}(\theta_e,\sigma_e^2)$ with $\theta_e$ being the exact value and $\sigma_e^2$ being the variance of $\tilde{\theta}$. We set $\sigma_e^2=10^{-1.5}$. The transmit covariance matrix is optimized to minimize the CRB for the inexact location, $\mathrm{CRB}_\theta(\tilde{\theta})$ in (\ref{CRB}), under the constraints in (P1), which is a convex problem.
\end{itemize}

\begin{figure} [t]
	\centering
	\includegraphics[width=3.5in]{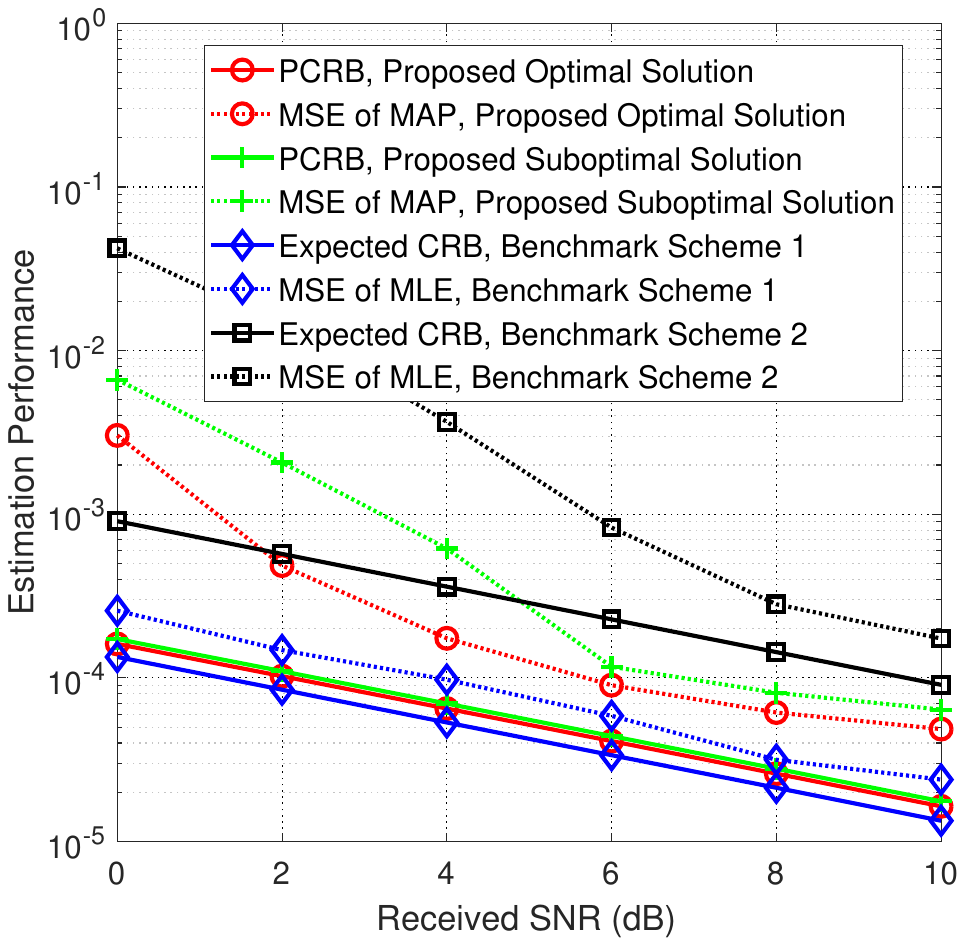}
	\vspace{-6mm}
	\caption{Sensing performance with different transmit covariance matrix designs.}\label{Fig_MCE2}
	\vspace{-6mm}	\end{figure}

In Fig. \ref{Fig_lowerbound1}, we evaluate the lower bound of MSE with different designs versus the communication rate target. For Benchmark Schemes 1 and 2, we adopt the expected CRB as the long-term MSE lower bound. It is observed that the PCRB achieved by the proposed suboptimal solution is close to that of the optimal solution in all feasible rate regimes; moreover, the PCRB upper bound is observed to be tight. This further verifies the effectiveness of using the proposed PCRB upper bound as a tractable sensing performance metric in the suboptimal solution. On the other hand, it is observed that the PCRB for our proposed solutions and the expected CRB for Benchmark Scheme 1 both increase with the communication rate target, since the limited power and spatial resources at the BS transmitter need to be allocated among the communication and sensing functions, thus leading to a trade-off.  When the rate target is low to moderate, the performance achieved by our proposed solutions is close to the expected CRB achieved by Benchmark Scheme 1 (which is a genie-aided scheme with additional exact target location information). When the rate target is high, the proposed solutions outperform Benchmark Scheme 1. This is because when only little resource is available for target sensing, the observation in $\bm{Y}$ carries little information due to the weak echo signals, while our proposed schemes can harvest extra information from the prior distribution, thus leading to an improved PCRB. Moreover, note that it is generally difficult to obtain the exact location of a sensing target before sending dedicated ISAC or sensing signals, while the prior distribution can be more easily obtained based on historic observations. Therefore, our proposed schemes are effective methods to design the transmit covariance matrix in practice. Finally, the expected CRB achieved by Benchmark Scheme 2 is much larger than the PCRB or expected CRB achieved by the other schemes. Particularly, it yields high expected CRB with low rate target since the transmitter resources are erroneously focused on the inexact location, and also with high rate target since the transmitter resources are focused on serving the communication user. The above results indicate that the sensing performance based on \emph{known} target location will be significantly degraded once the known location information is inexact.

Then, we show in Fig. \ref{Fig_MCE2} the actual sensing performance in terms of MSE achieved by different designs versus the received SNR at the BS receiver $\frac{P|\alpha|^2L}{\sigma_s^2}$. The transmitted signals in $\bm{X}$ are randomly generated according to the transmit covariance matrix $\bm{W}$. For the proposed solutions based on the prior PDF of $\theta$, the estimation of $\theta$ is obtained by the maximum a posteriori (MAP) estimation method, where $\hat{\theta}_{\mathrm{MAP}}=\arg\ \max\limits_\theta\;\left(  \ln f(\bm{Y}|\bm{\zeta})+\ln p_\Theta(\theta)\right)$ is obtained by searching over the region $[-\frac{\pi}{2},\frac{\pi}{2})$. For benchmark schemes based on a known angle, the maximum likelihood estimation (MLE) method is adopted to estimate $\theta$, where $\hat{\theta}_{\mathrm{MLE}}=\arg\ \max\limits_\theta\; \ln f(\bm{Y}|\bm{\zeta}) $ is obtained by searching over the region $[-\frac{\pi}{2},\frac{\pi}{2})$. It is observed from Fig. \ref{Fig_MCE2} that the MSE is bounded by the corresponding PCRB or expected CRB. Moreover, the proposed solutions outperform Benchmark Scheme 2, and the sensing performance achieved by the proposed solutions is close to that achieved by the genie-aided Benchmark Scheme 1 with a known exact location. This proves that the proposed solutions can achieve good sensing performance by exploiting prior distribution information.
	
\newcounter{mytempeqncnt}
\begin{figure*}[!t]
	\setcounter{mytempeqncnt}{\value{equation}}
	\begin{align}\label{inequality}
		& \left(\mathrm{tr}\left(\bm{A}_2\bm{W}\right) \mathrm{tr}\left(\bm{A}_4\bm{W}\right)- \left| \mathrm{tr}\left(\bm{A}_3\bm{W}\right)\right|^2 \right)/N_r^2 \nonumber\\
		=&\!\!\int \sum\limits_{i=1}^{N_t}u_i|\boldsymbol{\dot{a}}^H(\theta)\boldsymbol{w}_i|^2 p_\Theta(\theta)d\theta\int \sum\limits_{i=1}^{N_t}u_i|\boldsymbol{a}^H(\theta)\boldsymbol{w}_i|^2 p_\Theta(\theta)d\theta  -\left|\int \sum\limits_{i=1}^{N_t}u_i\boldsymbol{a}^H(\theta)\boldsymbol{w}_i\boldsymbol{w}_i^H\boldsymbol{\dot{a}}(\theta) p_\Theta(\theta)d\theta\right|^2 \nonumber\\
		=&\!\!\int\!\!\!\!\int\!\!\left(\!   \frac{1}{2}\!\left(\sum\limits_{i=1}^{N_t}u_i|\boldsymbol{\dot{a}}^H(\theta_a)\boldsymbol{w}_i|^2 \right)\!\!\left(\sum\limits_{i=1}^{N_t}u_i|\boldsymbol{ a}^H(\theta_b)\boldsymbol{w}_i|^2\right)  +\!\frac{1}{2}\!\left(\!\sum\limits_{i=1}^{N_t}\!u_i|\boldsymbol{ a}^H\!(\theta_a)\boldsymbol{w}_i|^2 \!\!\right)\!\!\left(\!\sum\limits_{i=1}^{N_t}\!u_i|\boldsymbol{\dot{a}}^H\!(\theta_b)\boldsymbol{w}_i|^2 \!\!\right)\! \right)p_\Theta(\theta_a)p_\Theta(\theta_b)d\theta_a d\theta_b \nonumber\\
		&\quad \!\!-\!\int\!\!\!\!\int\!\! \left( \sum\limits_{i=1}^{N_t}u_i\boldsymbol{a}^H(\theta_a)\boldsymbol{w}_i\boldsymbol{w}_i^H\boldsymbol{\dot{a}}(\theta_a) \! \right)\!\left(   \sum\limits_{i=1}^{N_t}u_i\boldsymbol{a}^H(\theta_b)\boldsymbol{w}_i\boldsymbol{w}_i^H\boldsymbol{\dot{a}}(\theta_b)\!  \right)  p_\Theta(\theta_a)p_\Theta(\theta_b)d\theta_a d\theta_b \nonumber\\
		=&\!\!\int\!\!\!\!\int\!\! \sum\limits_{n=1}^{N_t}\!\sum\limits_{i=1}^{N_t}\!\frac{u_nu_i}{2}\!\!\left(\! \left|\boldsymbol{\dot{a}}^H(\!\theta_a\!)\bm{w}_n\!\bm{w}_i^H\!\bm{a}(\!\theta_b\!)\right|^{\!2} \!\!-\!  2\!\left(\boldsymbol{\dot{a}}^H(\!\theta_a\!)\bm{w}_n\!\bm{w}_i^H\!\bm{a}(\!\theta_b\!)\right)\!\!\left(\boldsymbol{{a}}^H(\!\theta_a\!)\bm{w}_n\!\bm{w}_i^H\!\bm{\dot{a}}(\!\theta_b\!) \right) \!\!+\!\! \left|\boldsymbol{{a}}^H(\!\theta_a\!)\bm{w}_n\!\bm{w}_i^H\!\bm{\dot{a}}(\!\theta_b\!)\right|^{\!2}  \right)\! p_\Theta(\!\theta_a\!)p_\Theta(\!\theta_b\!)d\theta_a \!d\theta_b \nonumber\\
		=&\!\! \int\!\!\!\!\int\!\!\sum\limits_{n=1}^{N_t}\!\sum\limits_{i=1}^{N_t}\frac{u_nu_i}{2}\left|\boldsymbol{\dot{a}}^H(\theta_a)\bm{w}_n\bm{w}_i^H\bm{a}(\theta_b)  - \boldsymbol{{a}}^H(\theta_a)\bm{w}_n\bm{w}_i^H\bm{\dot{a}}(\theta_b)\right|^2p_\Theta(\theta_a)p_\Theta(\theta_b)d\theta_a d\theta_b \geq\! 0.
	\end{align}
	\hrulefill
	\vspace*{4pt}
\end{figure*}

\section{Conclusions}\label{sec_con}
This paper studied a MIMO ISAC system where the desired sensing parameter is unknown and random, with known prior distribution information. The PCRB of the parameter estimation MSE was first derived, for which a more tractable and tight upper bound was proposed. It was analytically shown that by exploiting the prior information, the PCRB is guaranteed to be no larger than the average CRB without exploiting the prior information. Next, the problem of transmit covariance matrix optimization was formulated to minimize the sensing PCRB, subject to a rate constraint at the communication user. The formulated problem was revealed to be equivalent to a convex problem, based on which the optimal solution was obtained, and useful properties on its rank were derived. A suboptimal solution in semi-closed form was then proposed by replacing the objective function with the PCRB upper bound, which requires lower complexity. The effectiveness of the proposed solutions was validated via extensive numerical results.

\appendices
\section{Proof of Proposition \ref{prop_bound}}\label{proof_bound}
First, we express $\bm{W}$ as $\bm{W}= \sum_{i=1}^{N_t}u_i\bm{w}_i\bm{w}_i^H$ where $u_i\geq0,\ i=1,...,N_t$  and $\bm{w}_i \in \mathbb{C}^{N_t\times 1},\ i=1,...,N_t$ denote the eigenvalues and eigenvectors in $\bm{W}$, respectively. Then, we have the inequality in (\ref{inequality}) at the top of this page. By applying (\ref{inequality}) to the denominator of (\ref{PCRB}), Proposition \ref{prop_bound} is proved.
\section{Proof of Proposition \ref{prop_rank}}\label{proof_rank}
In this case, the optimal solution to (P3) can be obtained by solving the following problem:
\begin{align}\label{P3_L}
	\mbox{(P3-II)} \quad \mathop{\mathrm{max}}_{\bm{W}\succeq \bm{0}} \;  \log_2\left|\!\bm{I}_{N_u}\!\!+\!\! \frac{\bm{H}\bm{W}\bm{H}^H}{\sigma_c^2}\!\right|\!\!-\!\!\mathrm{tr}\left(\! \hat{\bm{D}}\bm{W}\!\right),
\end{align}
where $\hat{\bm{D}}=\frac{\mu^\star_P}{\mu^\star_R}\bm{I}_{N_t}- \frac{1}{\mu^\star_R}\bm{D}^\star $.

Suppose $\mu_P\leq d_1$. Let $\bm{W}=\tilde{x}_W\bm{q}_1\bm{q}_1^H$
with $\tilde{x}_W$ being any positive constant denote a feasible solution to (P3-II). Substituting $\bm{W}$ into (P3-II) yields $\log_2\left(1+\frac{\tilde{x}_W\|\bm{H}\bm{q}_1\|^2}{\sigma_c^2}\right)+\frac{\tilde{x}_W}{\mu^\star_R}(d_1-\mu_P)$. Note that $\|\bm{H}\bm{q}_1\|>0$ is satisfied almost surely due to the linearly independent rows in $\bm{H}$. When $\tilde{x}_W \rightarrow \infty$, we have $ \log_2\left(1+\frac{\tilde{x}_W\|\bm{H}\bm{q}_1\|^2}{\sigma_c^2}\right)+\frac{\tilde{x}_W}{\mu^\star_R}(d_1-\mu_P) \rightarrow \infty$ and (P3-II) becomes unbounded. Thus, the assumption of $\mu_P \leq d_1$ is not true. Therefore, $\mu_P > d_1$ should be satisfied, which yields $\hat{\bm{D}}=\frac{\mu^\star_P}{\mu^\star_R}\bm{I}_{N_t}- \frac{1}{\mu^\star_R}\bm{D}^\star \succ \bm{0}$. Define $ \bm{M} =\hat{\bm{D}}^{ \frac{1}{2}}\bm{W}\hat{\bm{D}}^{\frac{1}{2}}\succeq \bm{0}$. Problem (P3-II) is equivalent to the following problem:
\begin{align}\label{P3_L2}
	\!\!\! \mbox{(P3-II')} \; \mathop{\mathrm{max}}_{  \bm{M} \succeq \bm{0}} \; \!\log_2\!\left|\bm{I}_{\!N_u}\!\!\!+\!\! \frac{\bm{H}\hat{\bm{D}}^{\!-\frac{1}{2}} \!\bm{M}\! \hat{\bm{D}}^{\!-\frac{1}{2}}\!\bm{H}^H}{\sigma_c^2}\right|\!\!-\!\!\mathrm{tr}\!\left(  \!\bm{M}\! \right).
\end{align}
Notice that Problem (P3-II') is a similar form as that of maximizing the Lagrangian of (P1-F), i.e., the MIMO channel rate maximization problem under a sum transmit power constraint \cite{Elements}. Based on the optimal solution structure given in Section \ref{sec_sol}, we have $\mathrm{rank}({\bm{M}}^\star)\leq\mathrm{rank}( \bm{H}\hat{\bm{D}}^{\!-\frac{1}{2}})=\min\left(N_t,\mathrm{rank}( \bm{H} ))=\mathrm{rank}( \bm{H} \right)$.

Note that for any feasible solution $ \bm{M} $ to (P3-II'), $\bm{W}=\hat{\bm{D}}^{-\frac{1}{2}} \bm{M} \hat{\bm{D}}^{-\frac{1}{2}}$ is a feasible solution to (P3-II) with the same objective value. Therefore, we have $\mathrm{rank}( {\bm{W}} ^\star) = \mathrm{rank}( \hat{\bm{D}}^{-\frac{1}{2}} \bm{M}^\star \hat{\bm{D}}^{-\frac{1}{2}})=\mathrm{rank}(\bm{M}^\star)\leq \mathrm{rank}( \bm{H} )$.

This thus completes the proof of Proposition \ref{prop_rank}.

\section{Proof of Proposition \ref{prop_solution}}\label{proof_solution}
To solve (P4) via the Lagrange duality method, we first solve problem $ \mathop{\mathrm{max}}\limits_{\bm{W} \succeq \bm{0}} \; \bar{\mathcal{L}}(\bm{W},\beta,\mu)$ with given dual variables $\beta$ and $\mu$, and then minimize the dual function $g(\beta,\mu)$ to find the optimal $\beta^\star$ and $\mu^\star$. Note that the solution to problem $\mathop{\mathrm{max}}\limits_{\bm{W} \succeq \bm{0}} \; \bar{\mathcal{L}}(\bm{W},\beta^\star,\mu^\star)$ is the optimal solution to (P4).

First, define $\bm{Q}=\frac{\mu}{\beta}\bm{I}_{N_t}- \frac{1}{\beta}\bm{A}_1$. Then, the problem $ \mathop{\mathrm{max}}\limits_{\bm{W} \succeq \bm{0}}\;  \bar{\mathcal{L}}(\bm{W},\beta,\mu)$ can be rewritten as
\begin{align}\label{Lagrangian2}
	\mbox{(P4-II)}\quad \mathop{\mathrm{max}}_{\bm{W}\succeq \bm{0}} \quad \log_2\left|\bm{I}_{N_u}+ \frac{\bm{H}\bm{W}\bm{H}^H}{\sigma_c^2}\right|-\mathrm{tr}\left( \bm{Q}\bm{W}\right).
\end{align}
Let $\lambda_1$ denote the largest eigenvalue of the matrix $\bm{A}_1$. Similar to the analysis in Appendix \ref{proof_rank}, to ensure that (P4-II) has a bounded optimal value, $\mu > \lambda_1$ should be satisfied. Consequently, $\bm{Q} \succ0$ holds and $\bm{Q}^{-1}$ exists. Define $\hat{\bm{W}}=\bm{Q}^{ \frac{1}{2}}\bm{W}\bm{Q}^{\frac{1}{2}}\succeq \bm{0}$. (P4-II) can be rewritten as
\begin{align}\label{Lagrangian3}
	\!\!\!\!\!\! \mbox{(P4-II')} \mathop{\mathrm{max}}_{\hat{\bm{W}}\succeq \bm{0}} \ \!\log_2\!\left|\bm{I}_{N_u}\!\!+\!\! \frac{\bm{H}\bm{Q}^{-\frac{1}{2}} \hat{\bm{W}}\bm{Q}^{-\frac{1}{2}}\!\bm{H}^H}{\sigma_c^2}\right|\!-\!\mathrm{tr}\left(  \!\hat{\bm{W}}\!\right).\!\!\!\!
\end{align}
Note that for any feasible solution $\hat{\bm{W}}$ to (P4-II'), $\bm{W}=\bm{Q}^{-\frac{1}{2}}\hat{\bm{W}}\bm{Q}^{-\frac{1}{2}}$ is a feasible solution to (P4-II) with the same objective value. Recall that the (reduced) SVD of $\bm{H}{\bm{Q}}^{-\frac{1}{2}}$ is given by $\bm{H}{\bm{Q}}^{-\frac{1}{2}}= \tilde{\bm{U}}\tilde{\bm{\Gamma}}^{\frac{1}{2}}\tilde{\bm{V}}^H$. Based on the KKT optimality conditions, the optimal solution to (P4-II') can be expressed as $\hat{\bm{W}}_U^\star=\tilde{\bm{V}}\tilde{\bm{\Lambda}} \tilde{\bm{V}}^H$, where $\tilde{\bm{\Lambda}} =\mathrm{diag}\{ \tilde{v}_1,..., \tilde{v}_{T}\}$, with $\tilde{v}_i\!=\!(1/\ln2\!-\!\sigma_c^2/\tilde{h}_i)^{+},\forall i$. The optimal solution to (P4-II) is then given by $\bm{W}_U^\star=\bm{Q}^{-\frac{1}{2}}\hat{\bm{W}}_U^\star\bm{Q}^{-\frac{1}{2}}=\bm{Q}^{-\frac{1}{2}}\tilde{\bm{V}}\tilde{\bm{\Lambda}} \tilde{\bm{V}}^H\bm{Q}^{-\frac{1}{2}}$.

On the other hand, to find the optimal $\beta$ and $\mu$, the dual problem $\mathop{\mathrm{min}}\limits_{\beta >0,\mu\geq0}\; g(\beta,\mu)$ can be solved by applying subgradient-based methods such as the ellipsoid method \cite{CVX22}. With given $\bm{W}^{\star}_U$, the subgradient of $g(\beta,\mu)$ at point $[\beta,\mu]$ is  given by $\left[  \log_2\left|\bm{I}_{N_u}+ \frac{\bm{H}\bm{W}_U^{\star}\bm{H}^H}{\sigma_c^2}\right|-\bar{R}, P-\mathrm{tr}\left( \bm{W}_U^{\star}\right) \right]$.

(P4) can be solved by iteratively updating $\bm{W}_U^{\star}$ and the dual variables $\beta$ and $\mu$. Specifically, the solution $\bm{W}_U^{\star}$ corresponding to the optimal dual variables $\beta^\star$ and $\mu^\star$ converges to the primal optimal solution to (P4). By substituting $\beta^\star$ and $\mu^\star$ into $\bm{W}_U^{\star}$, the proof of Proposition \ref{prop_solution} is completed.\footnote{We acknowledge \cite{SWIPT} for inspiring the proof of Proposition \ref{prop_solution}.}

\bibliographystyle{IEEEtran}
\bibliography{reference}

\begin{thebibliography}{10}
\providecommand{\url}[1]{#1}
\csname url@samestyle\endcsname
\providecommand{\newblock}{\relax}
\providecommand{\bibinfo}[2]{#2}
\providecommand{\BIBentrySTDinterwordspacing}{\spaceskip=0pt\relax}
\providecommand{\BIBentryALTinterwordstretchfactor}{4}
\providecommand{\BIBentryALTinterwordspacing}{\spaceskip=\fontdimen2\font plus
\BIBentryALTinterwordstretchfactor\fontdimen3\font minus
  \fontdimen4\font\relax}
\providecommand{\BIBforeignlanguage}[2]{{%
\expandafter\ifx\csname l@#1\endcsname\relax
\typeout{** WARNING: IEEEtran.bst: No hyphenation pattern has been}%
\typeout{** loaded for the language `#1'. Using the pattern for}%
\typeout{** the default language instead.}%
\else
\language=\csname l@#1\endcsname
\fi
#2}}
\providecommand{\BIBdecl}{\relax}
\BIBdecl

\bibitem{ISIT}
C.~Xu and S.~Zhang, ``{MIMO} radar transmit signal optimization for target
  localization exploiting prior information,'' in \emph{Proc. IEEE Int. Symp.
  Inf. Theory (ISIT)}, Jun. 2023, pp. 310--315.

\bibitem{6G}
D.~C. Nguyen, M.~Ding, P.~N. Pathirana, A.~Seneviratne, J.~Li, D.~Niyato,
  O.~Dobre, and H.~V. Poor, ``{6G} internet of things: {A} comprehensive
  survey,'' \emph{IEEE Internet Things J.}, vol.~9, no.~1, pp. 359--383, Jan.
  2022.

\bibitem{ISAC_survey}
A.~Liu, Z.~Huang, M.~Li, Y.~Wan, W.~Li, T.~X. Han, C.~Liu, R.~Du, D.~K.~P. Tan,
  J.~Lu, Y.~Shen, F.~Colone, and K.~Chetty, ``A survey on fundamental limits of
  integrated sensing and communication,'' \emph{IEEE Commun. Surv. Tut.},
  vol.~24, no.~2, pp. 994--1034, Feb. 2022.

\bibitem{ISAC_survey2}
F.~Liu, Y.~Cui, C.~Masouros, J.~Xu, T.~X. Han, Y.~C. Eldar, and S.~Buzzi,
  ``Integrated sensing and communications: {Toward} dual-functional wireless
  networks for {6G} and beyond,'' \emph{IEEE J. Sel. Areas Commun.}, vol.~40,
  no.~6, pp. 1728--1767, Jun. 2022.

\bibitem{signal_design1}
G.~N. Saddik, R.~S. Singh, and E.~R. Brown, ``Ultra-wideband multifunctional
  communications/radar system,'' \emph{IEEE Trans. Microw. Theory Technol.},
  vol.~55, no.~7, pp. 1431--1437, Jul. 2007.

\bibitem{signal_design2}
A.~Hassanien, M.~G. Amin, Y.~D. Zhang, and F.~Ahmad, ``Dual-function
  radar-communications: {Information} embedding using sidelobe control and
  waveform diversity,'' \emph{IEEE Trans. Signal Process.}, vol.~64, no.~8, pp.
  2168--2181, Apr. 2016.

\bibitem{signal_design3}
T.~Huang, N.~Shlezinger, X.~Xu, Y.~Liu, and Y.~C. Eldar, ``{MAJoRCom: A}
  dual-function radar communication system using index modulation,'' \emph{IEEE
  Trans. Signal Process.}, vol.~68, pp. 3423--3438, May 2020.

\bibitem{signal_design4}
D.~Garmatyuk, J.~Schuerger, K.~Kauffman, and S.~Spalding, ``Wideband {OFDM}
  system for radar and communications,'' in \emph{Proc. IEEE Radar Conf.},
  2009, pp. 1--6.

\bibitem{Shi}
Q.~Shi, L.~Liu, S.~Zhang, and S.~Cui, ``Device-free sensing in {OFDM} cellular
  network,'' \emph{IEEE J. Sel. Areas Commun.}, vol.~40, no.~6, pp. 1838--1853,
  Jun. 2022.

\bibitem{Liu}
L.~Liu and S.~Zhang, ``A two-stage radar sensing approach based on {MIMO-OFDM}
  technology,'' in \emph{Proc. IEEE Global Commun. Conf. (Globecom) Wkshps.},
  Dec. 2020.

\bibitem{Wang}
Q.~Wang, L.~Liu, S.~Zhang, and F.~Lau, ``Trilateration-based device-free
  sensing: {Two} base stations and one passive {IRS} are sufficient,'' in
  \emph{Proc. IEEE Global Commun. Conf. (Globecom)}, Dec. 2022.

\bibitem{Shi2}
Q.~Shi, L.~Liu, and S.~Zhang, ``Joint data association, {NLOS} mitigation, and
  clutter suppression for networked device-free sensing in {6G} cellular
  network,'' in \emph{Proc. IEEE Int. Conf. Acoustics Speech Signal Process.
  (ICASSP)}, Jun. 2023.

\bibitem{signal_design5}
R.~F. Tigrek, W.~J.~A. De~Heij, and P.~Van~Genderen, ``{OFDM} signals as the
  radar waveform to solve doppler ambiguity,'' \emph{IEEE Trans. Aerosp.
  Electron. Syst.}, vol.~48, no.~1, pp. 130--143, Jan. 2012.

\bibitem{jointDesign1}
F.~Liu, C.~Masouros, A.~Li, H.~Sun, and L.~Hanzo, ``{MU-MIMO} communications
  with {MIMO} radar: {From} co-existence to joint transmission,'' \emph{IEEE
  Trans. Wireless Commun.}, vol.~17, no.~4, pp. 2755--2770, Apr. 2018.

\bibitem{jointDesign2}
X.~Liu, T.~Huang, N.~Shlezinger, Y.~Liu, J.~Zhou, and Y.~C. Eldar, ``{Joint
  transmit beamforming for multiuser {MIMO} communications and {MIMO} radar},''
  \emph{IEEE Trans. Signal Process.}, vol.~68, pp. 3929--3944, Jun. 2020.

\bibitem{jointDesign3}
R.~Liu, M.~Li, Y.~Liu, Q.~Wu, and Q.~Liu, ``Joint transmit waveform and passive
  beamforming design for {RIS}-aided {DFRC} systems,'' \emph{IEEE J. Sel.
  Topics Signal Process.}, vol.~16, no.~5, pp. 995--1010, May 2022.

\bibitem{jointDesign4}
Z.~He, W.~Xu, H.~Shen, D.~W.~K. Ng, Y.~C. Eldar, and X.~You, ``Full-duplex
  communication for {ISAC: Joint} beamforming and power optimization,''
  \emph{IEEE J. Sel. Areas Commun.}, vol.~41, no.~9, pp. 2920--2936, Sep. 2023.

\bibitem{CRB1}
C.~R. Rao, ``Information and the accuracy attainable in the estimation of
  statistical parameters,'' in \emph{Breakthroughs in Statistics}.\hskip 1em
  plus 0.5em minus 0.4em\relax Springer, 1992, pp. 235--247.

\bibitem{CRB2}
I.~Bekkerman and J.~Tabrikian, ``{Target detection and localization using MIMO
  radars and sonars},'' \emph{IEEE Trans. Signal Process.}, vol.~54, no.~10,
  pp. 3873--3883, Sep. 2006.

\bibitem{CRB3}
R.~Boyer, ``{Performance bounds and angular resolution limit for the moving
  colocated MIMO radar},'' \emph{IEEE Trans. Signal Process.}, vol.~59, no.~4,
  pp. 1539--1552, Apr. 2011.

\bibitem{jointDesign5}
F.~Liu, Y.-F. Liu, A.~Li, C.~Masouros, and Y.~C. Eldar, ``{Cram\'{e}r-Rao bound
  optimization for joint radar-communication beamforming},'' \emph{IEEE Trans.
  Signal Process.}, vol.~70, pp. 240--253, Dec. 2022.

\bibitem{jointDesign6}
H.~Hua, T.~X. Han, and J.~Xu, ``{MIMO} integrated sensing and communication:
  {CRB-Rate} tradeoff,'' \emph{IEEE Trans. Wireless Commun.}, 2023, {Early
  Access}.

\bibitem{jointDesign7}
\BIBentryALTinterwordspacing
X.~Song, X.~Qin, J.~Xu, and R.~Zhang, ``{Cram\'{e}r-Rao} bound minimization for
  {IRS}-enabled multiuser integrated sensing and communications.'' [Online].
  Available: \url{https://arxiv.org/abs/2306.17493}
\BIBentrySTDinterwordspacing

\bibitem{PCRB1}
H.~L. Van~Trees, \emph{{Detection, Estimation, and Modulation Theory: Part I}},
  Wiley, New York, 1968.

\bibitem{Prior_CRB}
R.~Boyer and G.~Bouleux, ``Oblique projections for direction-of-arrival
  estimation with prior knowledge,'' \emph{IEEE Trans. Signal Process.},
  vol.~56, no.~4, pp. 1374--1387, Apr. 2008.

\bibitem{PCRB2}
P.~Tichavsky, ``Posterior {Cram\'{e}r-Rao} bound for adaptive harmonic
  retrieval,'' \emph{IEEE Trans. Signal Process.}, vol.~43, no.~5, pp.
  1299--1302, May 1995.

\bibitem{BCRB}
J.~Dauwels, ``Computing {Bayesian Cram\'{e}r-Rao} bounds,'' in \emph{Proc. IEEE
  Int. Symp. Inf. Theory (ISIT)}, Sep. 2005, pp. 425--429.

\bibitem{PCRB3}
P.~Tichavsky, C.~Muravchik, and A.~Nehorai, ``Posterior {Cram\'{e}r-Rao} bounds
  for discrete-time nonlinear filtering,'' \emph{IEEE Trans. Signal Process.},
  vol.~46, no.~5, pp. 1386--1396, May 1998.

\bibitem{PCRB4}
L.~Zuo, R.~Niu, and P.~K. Varshney, ``Conditional posterior {Cram\'{e}r-Rao}
  lower bounds for nonlinear sequential bayesian estimation,'' \emph{IEEE
  Trans. Signal Process.}, vol.~59, no.~1, pp. 1--14, Jan. 2011.

\bibitem{Active}
Z.~Zhang, T.~Jiang, and W.~Yu, ``Active sensing for localization with
  reconfigurable intelligent surface,'' in \emph{Proc. IEEE Int. Conf. Commun.
  (ICC)}, May 2023.

\bibitem{detection1}
W.~Huleihel, J.~Tabrikian, and R.~Shavit, ``Optimal adaptive waveform design
  for cognitive {MIMO} radar,'' \emph{IEEE Trans. Signal Process.}, vol.~61,
  no.~20, pp. 5075--5089, Jun. 2013.

\bibitem{detection2}
N.~Sharaga, J.~Tabrikian, and H.~Messer, ``Optimal cognitive beamforming for
  target tracking in {MIMO} radar/sonar,'' \emph{IEEE J. Sel. Topics Signal
  Process.}, vol.~9, no.~8, pp. 1440--1450, Dec. 2015.

\bibitem{ISAC_PCRB1}
K.~M. Attiah and W.~Yu, ``Active beamforming for integrated sensing and
  communication,'' in \emph{Proc. IEEE Int. Conf. Commun. (ICC) Wkshps.}, May
  2023.

\bibitem{ISAC_PCRB2}
B.~Teng, X.~Yuan, R.~Wang, and S.~Jin, ``Bayesian user localization and
  tracking for reconfigurable intelligent surface aided mimo systems,''
  \emph{IEEE J. Sel. Topics Signal Process.}, vol.~16, no.~5, pp. 1040--1054,
  Aug. 2022.

\bibitem{random2}
A.~W. Visser, ``Using random walk models to simulate the vertical distribution
  of particles in a turbulent water column,'' \emph{Mar. Ecol. Prog. Ser.},
  vol. 158, pp. 275--281, Nov. 1997.

\bibitem{SecureISAC}
K.~Hou and S.~Zhang, ``Secure integrated sensing and communication exploiting
  target location distribution,'' in \emph{Proc. IEEE Global Commun. Conf.
  (Globecom)}, Dec. 2023.

\bibitem{shen}
Y.~Shen and M.~Z. Win, ``{Fundamental limits of wideband localization---Part I:
  A general framework},'' \emph{IEEE Trans. Inf. Theory}, vol.~56, no.~10, pp.
  4956--4980, Oct. 2010.

\bibitem{sonars}
I.~Bekkerman and J.~Tabrikian, ``Target detection and localization using {MIMO}
  radars and sonars,'' \emph{IEEE Trans. Signal Process.}, vol.~54, no.~10, pp.
  3873--3883, Sep. 2006.

\bibitem{arxiv_gap}
\BIBentryALTinterwordspacing
X.~Gao, M.~Sitharam, and A.~E. Roitberg, ``Bounds on the {Jensen} gap, and
  implications for mean-concentrated distributions.'' [Online]. Available:
  \url{https://arxiv.org/abs/1712.05267}
\BIBentrySTDinterwordspacing

\bibitem{Elements}
T.~M. Cover and J.~A. Thomas, \emph{Elements of Information Theory}.\hskip 1em
  plus 0.5em minus 0.4em\relax New York: Wiley, 1991.

\bibitem{schur}
F.~Zhang, \emph{{The Schur Complement and Its Applications}}.\hskip 1em plus
  0.5em minus 0.4em\relax Springer Science \& Business Media, 2006, vol.~4.

\bibitem{CVX}
S.~Boyd and L.~Vandenberghe, \emph{Convex Optimization}.\hskip 1em plus 0.5em
  minus 0.4em\relax Cambridge, U.K.: Cambridge Univ. Press, 2004.

\bibitem{cvxtool}
\BIBentryALTinterwordspacing
M.~Grant and S.~Boyd. {(Jun. 2015). \textit{CVX: MATLAB Software for
  Disciplined Convex Programming}}. [Online]. Available:
  \url{http://cvxr.com/cvx/}
\BIBentrySTDinterwordspacing

\bibitem{Palomar}
Y.~Huang and D.~P. Palomar, ``Rank-constrained separable semidefinite
  programming with applications to optimal beamforming,'' \emph{IEEE Trans.
  Signal Process.}, vol.~58, no.~2, pp. 664--678, Feb. 2010.

\bibitem{EVD}
I.~Kodrasi and S.~Doclo, ``Analysis of eigenvalue decomposition-based late
  reverberation power spectral density estimation,'' \emph{IEEE/ACM Trans.
  Audio, Speech, Language Process.}, vol.~26, no.~6, pp. 1106--1118, Jun. 2018.

\bibitem{CVX22}
\BIBentryALTinterwordspacing
S.~Boyd, ``Convex optimization {II},'' Stanford University. [Online].
  Available: \url{http://www.stanford.edu/class/ee364b/lectures.html}
\BIBentrySTDinterwordspacing

\bibitem{SWIPT}
R.~Zhang and C.~K. Ho, ``{MIMO} broadcasting for simultaneous wireless
  information and power transfer,'' \emph{IEEE Trans. Wireless Commun.},
  vol.~12, no.~5, pp. 1989--2001, May 2013.

\end{thebibliography}
\end{document}